\documentclass[runningheads]{llncs}

\usepackage[noadjust]{cite}

\usepackage{amsmath,amssymb,amsfonts}
\usepackage{algorithmic}
\usepackage{graphicx}
\usepackage{tikz}
\usepackage{pgfplots}
\usepackage{stfloats}
\usepackage{float}
\usepackage{textcomp}
\usepackage{xcolor}
\usepackage[all]{xy} 
\usepackage{bm}
\usepackage{multirow}

\def\BibTeX{{\rm B\kern-.05em{\sc i\kern-.025em b}\kern-.08em
    T\kern-.1667em\lower.7ex\hbox{E}\kern-.125emX}}

\newcommand{\beq}{\begin{equation}}
\newcommand{\enq}{\end{equation}}
\newcommand{\bel}{\begin{lemma}}
\newcommand{\enl}{\end{lemma}}
\newcommand{\bet}{\begin{theorem}}
\newcommand{\ent}{\end{theorem}}

\newcommand*{\cC}{\Co}

\newcommand{\suppress}[1]{}

\mathchardef\mhyphen="2D

\def\sM{\mathsf{M}}
\def\sL{\mathsf{L}}
\def\sN{\mathsf{N}}

\makeatletter
\newcommand*{\rom}[1]{\expandafter\@slowromancap\romannumeral #1@}
\makeatother

\mathchardef\mhyphen="2D

\makeatletter

\newcommand{\Rmnum}[1]{\expandafter\@slowromancap\romannumeral #1@}
\makeatother

\newcommand{\FF}{\mathbb{F}}

\newcommand{\Co}{C}

\graphicspath{{./fig/}}
\DeclareGraphicsExtensions{.pdf} 

\def\QED{\mbox{\rule[0pt]{1.5ex}{1.5ex}}}

\def\Label#1{\label{#1}\ [\ \text{#1}\ ]\ }
\def\Label{\label}

\begin{document}

\title{Adaptive Coding for Two-Way Wiretap Channel under Strong Secrecy\thanks{Supported 
in part by the National Natural Science Foundation of China
under Grant 62171212.}}

\author{Yanling Chen\inst{1}\orcidID{0000-0003-1603-9121} and Masahito~Hayashi\inst{2,3,4}\orcidID{0000-0003-3104-1000}}

\authorrunning{Y. Chen and M. Hayashi}
%
\institute{Volkswagen Infotainment GmbH, Germany
\email{yanling.chen@volkswagen-infotainment.com}\and
School of Data Science, The Chinese University of Hong Kong, Shenzhen, Longgang
518172, China \and
Shenzhen International Quantum Academy (SIQA),
Futian, Shenzhen 518048, China \and
Graduate School
of Mathematics, Nagoya University, Chikusa-ku, Nagoya 464-8602, Japan\\
\email{hmasahito@cuhk.edu.cn}}

%
%
%
\maketitle

\begin{abstract}
This paper studies adaptive coding for the two-way wiretap channel. Especially, the strong secrecy metric is of our interest that is defined by the information leakage of transmitted messages to the eavesdropper. 
First, we consider an adaptive coding, the construction of which is based on running the well studied non-adaptive coding in several rounds and the dependency between the adjacent rounds of transmission is introduced by the key exchange mechanism that is embedded in the non-adaptive coding in each transmission round.  As a result, we analyze the reliability and strong secrecy that are measured by the decoding error probability and information leakage, characterize them in terms of the {\it conditional R\'enyi mutual information}, and derive inner bounds on the secrecy capacity regions for the TW-WC under strong joint and individual secrecy constraints. 
Second, we introduce another adaptive coding method that explores the correlation among the outputs at the receivers. With this approach, we show that for the two-way wiretap channel that fulfills the conditionally independent condition, positive transmission rates can be always guaranteed even under the joint secrecy constraint.
\end{abstract}
\section{Introduction}
The two-way communications channel (TWC) was first introduced by Shannon in his pioneering paper \cite{Shannon1961}, where he established an inner bound and an outer bound for the capacity region of a discrete memoryless TWC. Remarkably, Dueck \cite{Dueck1979} showed that the capacity region of the TWC can exceed Shannon's inner bound. More specifically, Schalkwijk \cite{Schalkwijk1983} proposed a constructive coding approach for the binary multiplier channel (BMC), demonstrating that an achievable rate pair that is beyond Shannon's inner bound is attainable. So far it is known that Shannon's inner bound does not coincide with Shannon's outer bound in general. Although the capacity region of the TWC has been determined for some special cases, as some examples can be seen in  \cite{Shannon1961, Han1984, Varshney2013, SAL2016, CVA2017}, determining the capacity region of a general TWC in terms of single-letter characterization still remains as an open research problem. 

Let us take a close look into the Shannon's inner bound and outer bound. It is easy to notice that the inner bound is achieved using non-adaptive encoding, where the channel input at each user is constructed from the message to be transmitted only (thus it is independent of the past received signals). However, the outer bound allows the channel inputs of the two users dependent, where the dependence could be introduced by the adaption of channel input to the past received signals at each user. It is worth mentioning that for those special cases of TWC, the capacity region of which has been determined, in most cases, the capacity region could be attained by using only non-adaptive codes. For those TWCs where the adaption is necessary from a capacity perspective, the question is how to introduce the dependency and what is the right amount of dependency? 

Speaking of the two-way wiretap channel (TW-WC), it was first considered in \cite{TY2007, TY2008} for both the Gaussian TW-WC and the binary additive TW-WC. 
Inner bounds on the secrecy capacity regions (subject to weak joint secrecy constraint \cite{Wyner1975}: $I(M_1, M_2;\mathbf{Z})/H(M_1, M_2) \to 0$)
for both channels were derived. Note that these achievable secret-rate regions were established by using cooperative jamming, which is kind of non-adaptive coding as it did not exploit the advantage of the previous received signals for the channel input. In fact, it was shown in \cite{HY2013} that adaptive coding could be highly beneficial for certain Gaussian TW-WCs by utilizing the previous received signals. 
Moreover, \cite{GKYG2013} considered the general discrete memoryless TW-WC channel, and derived an inner bound of the capacity region under the weak joint secrecy by combining cooperative jamming and a secret-key exchange mechanism (i.e., each legitimate user sacrifices part of its secret rate to transmit a key to the other legitimate user in each transmission round). Remarkably, for the Gaussian TW-WC, the region obtained in \cite{GKYG2013} was shown to be strictly larger than the regions given in \cite{TY2008}, demonstrating the effectiveness of the adaptive coding (as each legitimate user adapts the key received in the previous round to encrypt part of its message) in terms of enlarging the existing secret-rate region achieved by non-adaptive coding in \cite{TY2008}. Notably, there are also other secrecy measures for multi-user communications \cite{CKV18}.  Especially,  the inner bounds on the secrecy capacity regions for the TW-WCs under one-sided and individual secrecy constraints are established in \cite{QCHT2016} and \cite{QDT2017}, respectively.

Besides these studies on TW-WCs under weak secrecy, the strong joint secrecy constraint on the general discrete memoryless TW-WC was considered in \cite{PB2011}, where the information leakage to the eavesdropper (rather than the leakage rate) is required to be negligible (i.e., $I(M_1, M_2;\mathbf{Z}) \to 0$). Note that \cite{PB2011} not only provided the inner bounds on the secrecy capacity regions by using cooperation jamming only (i.e., non-adaptive coding), but also recovered the achievable secret-rate region obtained in \cite{GKYG2013} by using cooperative jamming and key exchange (i.e., adaptive coding as in \cite{GKYG2013}). 
Moreover, a generic adaptive coding strategy was discussed therein that assumed the existence of a discrete memory source (without exact characterization). 
Another paper addressed the TW-WC under strong secrecy is \cite{HC2023}, where the authors studied the TW-WC under a strong joint or individual secrecy constraint, with focus on non-adaptive coding. More specifically, the secrecy and error exponents while using the non-adaptive coding for the TW-WC were characterized by the conditional R\'enyi mutual information. 

In this paper, we focus on the adaptive coding for the TW-WC. First, we consider the adaptive coding that is proposed in \cite{GKYG2013} and extend the result of \cite{HC2023} on characterizing the secrecy and error exponents while using a non-adaptive coding to the case of employing an adaptive coding.  The work is based on the observations that the construction of the adaptive coding proposed in \cite{GKYG2013} can be considered as running several rounds of non-adaptive coding; and the dependency (between the adjacent transmission rounds) is introduced by the key exchange. As an byproduct, we also derive inner bounds on the strong secrecy capacity regions for the TW-WC under joint and individual secrecy constraints. Note that our joint secrecy region is the same as the weak secrecy region derived in \cite{GKYG2013} and strong secrecy region recovered in \cite{PB2011}, while we strengthen the strong secrecy result by the characterization of the error and secrecy exponents. On the other hand, our strong individual secrecy region result improves the weak secrecy result in \cite{QDT2017} in the manner that not only the secrecy is strengthened from weak to strong, but also the achievable rate region is potentially enlarged as a gain by using an adaptive coding approach. 
As the second part of our contribution, we introduce an adaptive coding method that explores the correlation among the outputs at the receivers. 
In particular, we consider the case when the channel outputs are conditionally independent, i.e., the channel outputs are independent when the inputs are given.
We show that for the two-way wiretap channel that fulfills this conditionally independent condition, positive transmission rates can be always guaranteed even under the joint secrecy constraint, when we allow such an adaptive coding.

The rest of the paper is organized as follows. In Section \ref{sec:model}, we describe the system model, give definitions of a non-adaptive or adaptive code, joint and individual secrecy constraints and present our results on the attained inner bounds. 
In Section \ref{sec: Renyi Lemmas}, we give the definitions of  {R\'enyi mutual information} and  {conditional R\'enyi mutual information} that are used to characterize the secrecy and error exponents; we also recall the reliability and resolvability lemmas that are used in reliability and secrecy analysis. 
In Section \ref{sec: Non-Adaptive Coding}, we review the non-adaptive coding and the corresponding strong joint and individual secrecy results.   
In Section \ref{sec: Adaptive coding}, we consider the adaptive coding proposed in \cite{GKYG2013},  characterize both the secrecy and error exponents, and derive our inner bounds on the capacity regions for the TW-WC under strong joint and individual secrecy constraint. 
In Section \ref{sec: Exploring the correlation}, we discuss a different adaptive coding method that explores the correlation among the outputs at the receivers, showing the effectiveness of such a method by studying the case when the channel outputs are conditionally independent for given inputs.
We conclude the paper in Section \ref{Sec: Conclusion}.

\section{Channel model and strong secrecy result}\Label{sec:model}
\subsection{Formulation}
We consider a discrete memoryless two-way wiretap channel, where 
two legitimate users, User 1 and User 2 intend to exchange 
messages with each other in the presence of an external eavesdropper.
The channel is characterized by $P_{Y_1,Y_2,Z|X_1,X_2}.$ 
In this model, User $i$ has the input symbol set ${\cal X}_i$ and 
the output symbol set ${\cal Y}_i,$ where $i=1,2$.
The external eavesdropper has the output symbol set ${\cal Z}$.
The channel model is shown in Fig. \ref{fig: TW-WTC with an external eavesdropper}. 

\begin{figure}[htbp]
	\centering
	\includegraphics[width=.75\textwidth]{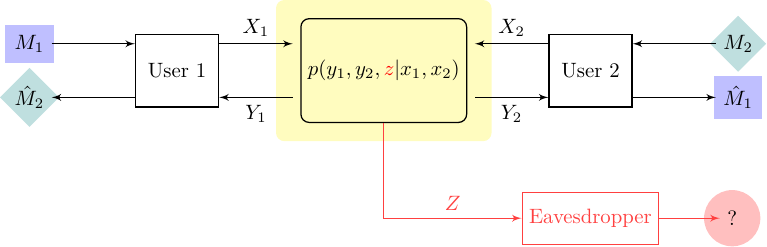}
	\caption{Two-Way wiretap channel with an external eavesdropper.} \Label{fig: TW-WTC with an external eavesdropper}
\end{figure}

Their messages $M_j$ are assumed to be uniformly distributed over  the message sets
$\mathcal{M}_i=[1:2^{nR_i}]$ for $i=1,2$.
For their communication, they use the above channel $n$ times,
and 
we denote User $i$'s channel input and output 
by $X_i^n=(X_{i,1}, \ldots, X_{i,n}) \in {\cal X}_i^n$ and 
$Y_i^n=(Y_{i,1}, \ldots, Y_{i,n})\in {\cal Y}_i^n$, respectively.
Also, we denote the channel output at the eavesdropper by 
$Z^n=(Z_{1}, \ldots, Z_{n})\in {\cal Z}^n$. 

We consider two types of encoders for two users.
\begin{itemize}
	\item The first type of encoder is a {\it non-adaptive} encoder $\phi_i$, which stochastically assigns the whole input $X_i^n$ based on the message $M_i$  for $i=1,2$.
	\item The second type of encoder is an {\it adaptive} encoder $\phi_i$, which	stochastically assigns the $t$-th input $X_{i,t}$ based on the message $M_i$ and 
	the previous outputs $Y_{i,1},\ldots, Y_{i,t-1}$ for $t=1,\ldots,n$ and $i=1,2$.
\end{itemize}

The decoder $\psi_i$ of User $i$ is given as a map from $M_i,$ ${\cal X}_i^n$, ${\cal Y}_i^n$ to ${\cal M}_{i \oplus 1}.$ 

A $(e^{nR_1}, e^{nR_2}, n)$ secrecy code 
$\Co_n$ for the two-way wiretap channel consists of
$2$ message sets $\mathcal{M}_1, \mathcal{M}_2$,
$2$ encoders $\phi_1$, $\phi_2$, and
$2$ decoders $\psi_1$, $\psi_2$. Whether the code is adaptive or non-adaptive depends on whether adaptive or non-adaptive encoders are used. 

To evaluate the {\it reliability} of the transmission, we consider the \emph{average probability of decoding error} at the legitimate receiver that is defined by
{
\begin{equation}\Label{eqn: Pe}
P_{e}^n(\Co_n)=\frac{1}{2^{n[R_1+R_2]}}\sum_{m_1, m_2}\Pr\left\{\bigcup\limits_{i\in \{1,2\}} \{m_i\neq \hat{m}_i\}\Bigg|\Co_n\right\}.
\end{equation}
}
 
To evaluate the {\it secrecy} of the transmission, we consider the strong secrecy that measures the amount of information (of $(M_1, M_2)$ jointly or of $M_1, M_2$ individually) leaked to the eavesdropper (through his/her observation $Z^n$).
We say that the rate pair $(R_{1,n},R_{2,n})$ is \emph{achievable under the strong joint secrecy constraint by adaptive (non-adaptive) codes}, if there exists a sequence of $(e^{nR_{1,n}}, e^{nR_{2,n}}, n)$ adaptive (non-adaptive) codes $\{\Co_n\}$ such that
$R_{i,n} \to R_i$ for $i=1,2$ and 
	\begin{align} 
	  P_{e}^n(\Co_n) &\leq \epsilon_n, \Label{eq:Reliability} \\
	  I(M_1, M_2;Z^n|\Co_n)	&\leq	\tau_n, \Label{eq:SSec}
\end{align}
with $\lim\limits_{n\to\infty} \epsilon_n = 0$ and $\lim\limits_{n\to\infty} \tau_n= 0.$ 
We say that the rate pair $(R_1,R_2)$ is \emph{achievable under the strong individual secrecy constraint by adaptive (non-adaptive) codes}, 
if there exists a sequence of $(e^{nR_{1,n}}, e^{nR_{2,n}}, n)$ adaptive (non-adaptive) codes $\{\Co_n\}$ such that
$R_{i,n} \to R_i$ for $i=1,2$ and 
	\begin{align} 
	  P_{e}^n(\Co_n) &\leq \epsilon_n, \Label{eq:Reliability2} \\
	  I(M_1;Z^n|\Co_n)	&\leq	\tau_n, \Label{eq:SSec2} \\
	  I( M_2;Z^n|\Co_n)	&\leq	\tau_n. \Label{eq:SSec3}
\end{align}

We define the achievability under the \emph{weak} joint secrecy constraint by adaptive (non-adaptive) codes by replacing $\tau_n$ by $n \tau_n$ in \eqref{eq:SSec}.
In the same way, we define the achievability under the \emph{weak} 
individual secrecy constraint by adaptive (non-adaptive) codes
by replacing $\tau_n$ by $n \tau_n$ in \eqref{eq:SSec2} and \eqref{eq:SSec3}.

\subsection{Achievable strong secrecy regions}\label{2B}

To consider the strong joint secrecy criteria \eqref{eq:SSec}, we define for a joint distribution $P_{V_1,V_2,X_1,X_2}$ the region
\begin{align}
&	{\cal R}_{J,A}(P_{V_1,V_2,X_1,X_2})\notag\\
:= &
	\left\{(R_1,R_2) \in \mathbb{R}_+^2
	\left|	
	\begin{array}{l}
		R_1 \leq I(Y_2;V_1|X_2), \\
		R_2 \leq I(Y_1;V_2|X_1), \\
		R_1+R_2  \leq  I(Y_2;V_1|X_2) + I(Y_1;V_2|X_1) - I(V_1, V_2;Z) 
	\end{array}
	\right.\right\}
	\Label{BPW1A}
\end{align}
where
$\mathbb{R}_+$ is the set of positive real numbers.\par

To consider the strong individual secrecy criteria \eqref{eq:SSec2} and \eqref{eq:SSec3}, we define for a joint distribution $P_{V_1,V_2,X_1,X_2}$ the region
\begin{align}
&	{\cal R}_{I, A}(P_{V_1,V_2,X_1,X_2})\notag\\
:= &
	\left\{(R_1,R_2)\in\mathbb{R}_+^2
	\left|
	\begin{array}{l}
		R_1 \leq I(Y_2;V_1|X_2),\\
		R_2 \leq I(Y_1;V_2|X_1),\\
		\max\{R_1, R_2\} \\
		\leq  I(Y_2;V_1|X_2) + I(Y_1;V_2|X_1) - I(V_1, V_2;Z)\\
		R_1 + R_2 \\
		\leq 
		\footnotesize
		{
		\left\{
		\begin{array}{c}
			\min\left\{
			\begin{array}{l}
				I(Y_2;V_1|X_2)-I(Z;V_1),\\
				I(Y_2;V_1|X_2) + I(Y_1;V_2|X_1) - I(V_1, V_2;Z)
			\end{array}
			 \right\}\\
			+ \\
			\min\left\{
			\begin{array}{l}
				I(Y_1;V_2|X_1)-I(Z;V_2),\\
				I(Y_2;V_1|X_2) + I(Y_1;V_2|X_1) - I(V_1, V_2;Z)
			\end{array}
			\right\}\\
		\end{array}
		\right\}
		}
	\end{array}
	\right.\right\}\Label{BPW2A}
\end{align}

We also define 
\begin{equation}\label{def: Q}
	{\cal Q}:=\{\prod\limits_{i\in \{1,2\}}P_{V_i}(v_i)P_{X_i|V_i}(x_i|v_i)\}
\end{equation}
and the following regions: 
\begin{align}
{\cal R}_{J,A}:=& cl. \bigcup_{P_{V_1,V_2,X_1,X_2} \in {\cal Q}}{\cal R}_{J,A}(P_{V_1,V_2,X_1,X_2}), \label{eqn: RJA}\\
{\cal R}_{I,A}:= &cl. \bigcup_{P_{V_1,V_2,X_1,X_2}\in {\cal Q}}{\cal R}_{I,A}(P_{V_1,V_2,X_1,X_2}), \label{eqn: RIA}
\end{align}
where $cl.$ expresses the closure of the convex hull.

We have the following theorem for the achievable secrecy rate region of the two-way wiretap channel.
\begin{theorem}\Label{Cor: S-MAC-KS} 
${\cal R}_{J,A}$ and ${\cal R}_{I,A}$ are achievable by adaptive codes under strong joint secrecy and strong individual secrecy, respectively.
\end{theorem} 

\section{Preliminaries}\Label{sec: Renyi Lemmas}
In this section we introduce some definitions and lemmas that will be used in the paper.
\subsection{Conditional R\'enyi mutual information}
We recall the {\it R\'enyi relative entropy} as
\begin{align}\Label{eqn: Renyi relative entropy}
	D_{1+s}(P \| Q):=\frac{1}{s}\log  \sum_{x}P(x)^{1+s} Q(x)^{-s}.
\end{align}
Note that $D_{1+s}(P \| Q)$ is nondecreasing w.r.t. $s$ for $s>0$ and $\lim\limits_{s\to 0}D_{1+s}(P \| Q)=D(P \| Q),$ i.e., the relative entropy.

Following the notations in \cite[Eq. (36)]{HT16},
\cite[Eq. (4.14)]{Springer}, and
\cite[Eqs. (50), (52), and (58)]{TH},
we define the {\it R\'enyi mutual information} as
\begin{align}
	I_{1+s}^{\uparrow} ( Z ; X) 
	:=
	D_{1+s} (P_{Z X}
	\| P_Z \times P_{X}   ).\Label{Ne1}
\end{align}
and
the {\it conditional R\'enyi mutual information} as
\begin{align}
	e^{-s I_{\frac{1}{1+s}}^{\downarrow} ( Z ; X|Y) }
	:=
	\sum_y P_Y(y)
	e^{-s \min\limits_{Q_{Z|Y=y}}D_{\frac{1}{1+s}} (P_{Z X|Y=y} \|  Q_{Z|Y=y} \times P_{X|Y=y}   )}.
\end{align}
The minimizer $Q_{Z|Y}^*$ is given as
\begin{equation}
	Q_{Z|Y}^*(z|y)= \frac{\sum\limits_{ x} P_{X|Y}(x|y)
		P_{Z| X Y}(z|x,y)^{\frac{1}{1+s}}}{\sum\limits_{z'} \sum\limits_{ x'} P_{X|Y}(x'|y)
		P_{Z| X Y}(z'|x',y)^{\frac{1}{1+s}}},
\end{equation}
and $I_{\frac{1}{1+s}}^{\downarrow} ( Z ; X| Y )$
is written in the following form \cite[Eq. (54)]{TH};
\begin{align}
&	e^{-s I_{\frac{1}{1+s}}^{\downarrow} ( Z ; X| Y ) }\notag\\
	=
	&\sum_{ y} P_Y(y)
	\sum_{z} P_{Z|Y}(z|y)\cdot\Big(
	\sum_{ x} P_{X|Y}(x|y)^{\frac{s}{1+s}}
	P_{X| Z Y}(x|z,y)^{\frac{1}{1+s}}
	\Big)^{1+s}  \nonumber\\
	=
	&\sum_{ y} P_Y(y)
	\sum_{z} 
	\Big(\!
	\sum_{ x} P_{X|Y}(x|y)
	\Big(\frac{P_{ZX| Y}(z,x|y)}{ P_{X| Y}(x|y)}\!\Big)^{\frac{1}{1+s}}
	\!
	\Big)^{1+s}  \nonumber\\
	=
	&\sum_{ y} P_Y(y)
	\sum_{z} 
	\Big(
	\sum_{ x} P_{X|Y}(x|y)
	P_{Z| X Y}(z|x,y)^{\frac{1}{1+s}}
	\Big)^{1+s} \Label{Ne3} .
\end{align}
When $X$ and $Y$ are independent of each other, 
it is simplified as
\begin{align}
	e^{-s I_{\frac{1}{1+s}}^{\downarrow} ( Z ; X| Y ) }
	=
	&\sum_{ y} P_Y(y)
	\sum_{z} 
	\Big(
	\sum_{ x} P_{X}(x)
	P_{Z| X Y}(z|x,y)^{\frac{1}{1+s}}
	\Big)^{1+s}  \nonumber \\
	=
	&\sum_{ y,z} 
	\Big(
	\sum_{ x} P_{X}(x)
	(P_Y(y) P_{Z| X Y}(z|x,y))^{\frac{1}{1+s}}
	\Big)^{1+s} \nonumber \\
	=&e^{-s I_{\frac{1}{1+s}}^{\downarrow} ( Z Y; X ) }
	\Label{Ne3Y} \\
	=&
	e^{-s \min\limits_{Q_{ZY}}D_{\frac{1}{1+s}} (P_{Z YX} \|  Q_{ZY} \times P_{X}   )}.
	\Label{Ne3G} 
\end{align}
In addition, the minimizer $Q_{ZY}^*$ in \eqref{Ne3G} is given as
\begin{equation}
	Q_{ZY}^*(z,y)=
	\frac{\sum\limits_{ x} P_{X}(x)
		P_{ZY| X }(z,y|x)^{\frac{1}{1+s}}}{\sum\limits_{z',y'} \sum\limits_{ x'} P_{X}(x')
		P_{ZY| X }(z',y'|x')^{\frac{1}{1+s}}}.
\end{equation}

\subsection{Resolvability and reliability lemmas}
We have the following resolvability lemma for the single shot analysis.
\begin{lemma}\cite[Lemma 2]{HC2023} \Label{L2} 
	Let $P_{Z|X_1 X_2}$ be a MAC with two input systems.
	Assume that $P_{X_1 X_2}=P_{X_1} \times P_{X_2}$.
	Let $X_{i,1},\ldots, X_{i,\sL_i}$ be the random variables that are independently generated subject to $P_{X_i},$ for $i=1,2$.
	Here, we denote this code selection by ${\cC}$ and define
	the distribution $P_{Z| \cC}$ for the variable $Z$ dependently of 
	the code selection ${\cC}$ as
	$
	P_{Z| \cC}(z):=
	\frac{1}{\sL_1 \sL_2} 
	\sum\limits_{j_1,j_2} P_{Z| X_1=X_{1,j_1}, X_2=X_{2,j_2}  }(z).
	$
	Then, for $s \in [0,1]$, we have $ 
	D(  P_{Z| \cC}\| P_{Z}) 
	\le  
	D_{1+s}(  P_{Z| \cC}\| P_{Z})$ and
	\begin{align*}
		s \mathbb{E}_{\cC} 
		\left[D_{1+s}(  P_{Z| \cC}\| P_{Z}) \right]
		\le 
		\sum_{\mathcal{S}\neq \emptyset, \mathcal{S}\subset \{1,2\}}
		\frac{1}{ \prod_{i \in \mathcal{S}} \sL_i^s} 
		e^{s I^{\uparrow}_{1+s} ( Z ; X_{\mathcal{S}}) }
	\end{align*}
	for $s \in [0,1]$.
\end{lemma}
Note that this lemma can be regarded as an
extension of the resolvability for the 2-transmitter MAC in \cite{src:Steinberg98}
by further characterizing the secrecy exponent of the coding mechanism. The extension of this lemma to $k$-transmitter MAC is given in the reference \cite[Lemma 1]{HC2019}.

As a reliability lemma, we have the following as a generalization of Gallager's bound \cite{Gallager68} for two-transmitter MAC.
\begin{lemma}\cite[Lemma 3]{HC2023} \Label{L3} 
	Let $P_{Y|X_1 X_2}$ be a MAC with two input systems.
	Assume that the $\sN$ messages are randomly selected as 
	the random variables
	$X_{2,1},\ldots, X_{2,\sN}$ that are independently generated subject to $P_{X_2}$.
	Also, assume that the receiver knows the other random variable $X_1$
	and selects it subject to $P_{X_1}$ independently of
	$X_{2,1},\ldots, X_{2,\sN}$.
	Here, we denote the above selection by ${\cC'}$ and 
	the decoding error probability under the maximum likelihood (ML) decoder
	by $e_{\cC'}$.
	Then, for $s \in [0,1]$, we have
	\begin{align}
		\mathbb{E}_{\cC'} 
		\left[e_{\cC'}\right]
		\le  
		\sN^s
		e^{-s I_{\frac{1}{1+s}}^{\downarrow} (Y X_{1};  X_{2})}
		= \sN^s
		e^{-s I_{\frac{1}{1+s}}^{\downarrow} (Y ;  X_{2}| X_{1})}.\Label{NMD}
	\end{align}
\end{lemma}

\section{Review on Non-Adaptive Coding}\label{sec: Non-Adaptive Coding}

In this section, we review the case where non-adaptive codes are used by both legitimate users. 
\subsection{Code construction}
\begin{figure}[h]
	\centering	
	\begin{tabular}{rcl}		
		$V_1^n:$ &	& 
		$\overbrace{	
			\begin{tikzpicture}
				\node[minimum height=1.6em, minimum width=6.5em, anchor=base, fill=blue!25] {$m_{1}$}; 
			\end{tikzpicture}
		}^{nR_{1}}
		\overbrace{
			\begin{tikzpicture}
				\node[minimum height=1.6em, minimum width=3.5em, anchor=base, fill=red!25] {$m_{1,r}$}; 
			\end{tikzpicture}
		}^{nR_{1,r}}$ \\
		$V_2^n:$ & 	& 
		$\underbrace{
			\begin{tikzpicture}
				\node[minimum height=1.6em,minimum width=7em, anchor=base, fill=teal!25] {$m_{2}$}; 
			\end{tikzpicture}
		}_{nR_2}$
	$\underbrace{
		\begin{tikzpicture}
			\node[minimum height=1.6em, minimum width=4em, anchor=base, fill=red!25] {$m_{2,r}$}; 
		\end{tikzpicture}
	}_{nR_{2, r}}$ 
	\end{tabular}
	\caption{Encoding for Non-Adaptive Coding.}
	\label{fig: Message structure a}	
\end{figure}
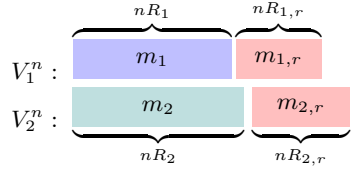
{\it Codebook Generation:}
At transmitter $i,$ let $V_{i,1}^n,\ldots, V_{i, \sM_i \cdot \sL_i}^n$ 
be random variables that are independently generated subject to 
$P_{V_i^n}=\prod_{j=1}^{n}P_{V_{i,j}}, $ where 
$V_i^n=(V_{i,1}, \cdots, V_{i,n}),$  
$\sM_i= e^{n R_i}$ and 
$\sL_i= e^{n R_{i,r}}$ for $i=1,2.$

\noindent {\it Encoding:}
When the legitimate user $i$ intends to send the message $m_i$, he(she) chooses one of 
${V}^n_{i,(m_i-1) \sL_i+1}, \ldots, {V}^n_{i,m_i \sL_i}$ with equal probability, which is denoted to be ${V}^n_{i,(m_i, m_{i,r})}$ and sends it. 

\noindent {\it Decoding:}
At the other legitimate receiver, an ML decoder will be used to decode ${V}^n_{i,(m_i, m_{i,r})}$ (and thus $m_{i}$)
by using the receiver's information $X_{i \oplus 1}^n$. 

\subsection{Exponential evaluation}
We have the following Lemma from \cite{HC2023} for the non-adaptive coding, which gives exponential evaluation on both the decoding error probability  and the information leaked to the eavesdropper.
\begin{lemma}\cite[Lemma 4]{HC2023} \Label{L3T}
	For a joint distribution $P_{V_1 V_2 X_1 X_2} \in {\cal Q}$,
	a rate pair $(R_1,R_2)$, and two positive numbers $R_{1,r},R_{2,r}$,
	there exists a sequence of $(e^{nR_1}, e^{nR_2}, n)$  non-adaptive codes $\Co_n$ such that
	\begin{align} 
		&P_{e}^n(\Co_n) \notag \\ \leq& \ 2\left[
		e^{ns\left(  R_1+R_{1,r}-  I_{\frac{1}{1+s}}^{\downarrow} (Y_2 ;  V_{1}| X_{2})\right)}
		+e^{ns\left(  R_2+R_{2,r}-  I_{\frac{1}{1+s}}^{\downarrow} (Y_1 ;  V_{2}| X_{1})\right)}\right]
		, \Label{eq:ReliabilityD} \\
		&I(M_1, M_2;Z^n|\Co_n)	\notag \\ \leq &	2\left[
		\sum_{\mathcal{S}\neq \emptyset, \mathcal{S}\subset  \{1,2\}}
		e^{n s\left( I^{\uparrow}_{1+s} ( Z ; V_{\mathcal{S}})-\sum\limits_{j \in \mathcal{S}}R_{j,r}\right) }\right]. \Label{eq:SSecD}
	\end{align}
	Also, there exists a sequence of $(e^{nR_1}, e^{nR_2}, n)$  non-adaptive codes $\Co_n$ such that
	\begin{align} 
&		P_{e}^n(\Co_n)\notag \\ 
\leq & 3 \left[e^{ns\left(  R_1+R_{1,r}-  I_{\frac{1}{1+s}}^{\downarrow} (Y_2 ;  V_{1}| X_{2})\right)}
		+e^{ns\left(  R_2+R_{2,r}-  I_{\frac{1}{1+s}}^{\downarrow} (Y_1 ;  V_{2}| X_{1})\right)}\right]
		, \Label{eq:Reliability2D} \\
		&I(M_1;Z^n|\Co_n)	\notag \\
		\leq &3 \Big[e^{n s\left( I^{\uparrow}_{1+s} ( Z ; V_1,V_2)-(R_{1,r}+R_{2,r}+R_2)\right) }
		+e^{n s\left( I^{\uparrow}_{1+s} ( Z ; V_1)- R_{1,r}\right) }	\notag \\
		&+e^{n s\left( I^{\uparrow}_{1+s} ( Z ; V_2)-(R_{2,r}+R_2)\right) }\Big]
		, \Label{eq:SSec2D} \\
		&I( M_2;Z^n|\Co_n)	\notag \\
		\leq & 3\Big[
		e^{n s\left( I^{\uparrow}_{1+s} ( Z ; V_1,V_2)-(R_{1,r}+R_{2,r}+R_1)\right) }
		+e^{n s\left( I^{\uparrow}_{1+s} ( Z ; V_2)- R_{2,r}\right) }
		\notag \\
		&+e^{n s\left( I^{\uparrow}_{1+s} ( Z ; V_1)-(R_{1,r}+R_1)\right) }\Big]
		. \Label{eq:SSec3D}
	\end{align}
\end{lemma}
\subsection{Achievable secrecy regions}
Considering the strong joint secrecy criteria \eqref{eq:SSec}, we define for a joint distribution $P_{V_1 V_2 X_1 X_2}$ the region
\begin{align}
&	{\cal R}_{J,N}(P_{V_1 V_2 X_1 X_2})\notag \\
:=& 
	\left\{(R_1,R_2) \in \mathbb{R}_+^2
	\left|	
	\begin{array}{l}
		R_1 \leq I(Y_2;V_1|X_2)-I(Z;V_1), \\
		R_2 \leq I(Y_1;V_2|X_1)-I(Z;V_2), \\
		R_1+R_2  \leq  I(Y_2;V_1|X_2) + I(Y_1;V_2|X_1) - I(V_1, V_2;Z) 
	\end{array}
	\right.\right\}
	\Label{BPW1}
\end{align}
where $\mathbb{R}_+$ is the set of positive real numbers.\par

Considering the strong individual secrecy criteria \eqref{eq:SSec2} and \eqref{eq:SSec3}, we define for a joint distribution $P_{V_1 V_2 X_1 X_2}$ the region
\begin{align}
&	{\cal R}_{I, N}(P_{V_1 V_2 X_1 X_2})\notag \\
:=& 
	\left\{(R_1,R_2)\in\mathbb{R}_+^2
	\left|
	\begin{array}{l}
		R_1 \leq I(Y_2;V_1|X_2)-I(Z;V_1),\\
		R_2 \leq I(Y_1;V_2|X_1)-I(Z;V_2),\\
		\max\{R_1, R_2\} \\
		\leq I(Y_1;V_2|X_1)+I(Y_2;V_1|X_2)-I(V_1, V_2;Z)
	\end{array}
	\right.\right\}\Label{BPW2}
\end{align}
Then, the region ${\cal R}_{J,N}$ and ${\cal R}_{I,N}$ are achievable by using non-adaptive codes, where 
	\begin{align}
		{\cal R}_{J,N}:=& cl. \bigcup_{P_{V_1 V_2 X_1 X_2} \in {\cal Q}}{\cal R}_{J, N}(P_{V_1 V_2 X_1 X_2}), \label{eqn: RJN}\\
		{\cal R}_{I,N}:= &cl. \bigcup_{P_{V_1 V_2 X_1 X_2}\in {\cal Q}}{\cal R}_{I,N}(P_{V_1 V_2 X_1 X_2}), \label{eqn: RIN}
	\end{align}
$cl.$ expresses the closure of the convex hull and ${\cal Q}$ is as defined in \eqref{def: Q}.

\section{Adaptive coding} \Label{sec: Adaptive coding}
In this section, we consider the case where both legitimate users use adaptive codes that were originally proposed in \cite{GKYG2013}. We provide exponential evaluation on both the decoding error probability and the information leaked to the eavesdropper as an extension of \cite{HC2023} for the non-adaptive codes. Note that our characterization of the error and secrecy exponents in Lemma \ref{L3TA} together with the reliability analysis in Section \ref{sec: reliability A} and strong secrecy analysis  in Section \ref{sec: Secrecy A} prove the validity of Theorem \ref{Cor: S-MAC-KS}. 

\subsection{Code Construction}
Recall that in the non-adaptive coding as described in Section \ref{sec: Non-Adaptive Coding}, to send $m_i,$ User $i$ sends ${V}^n_{i,(m_i, m_{i,r})}$ for $i=1,2.$ At the legitimate decoder, $(m_i, m_{i,r})$ will be decoded; whilst only secrecy of $m_i$ is required to be guaranteed from the eavesdropper. In other words, $m_i$ is the main message part to be kept secret; whilst  $m_{i,r}$ is the auxiliary message part that could be public.  

The adaptive coding proposed in \cite{GKYG2013} is based on the non-adaptive coding while running the non-adaptive code in several rounds. The dependency among the transmission rounds are introduced by the key exchange. In more detail, key exchange occurs at each transmission round and the exchanged key will be used in the next transmission round to encrypt partial message in a one-time-pad  manner \cite{Shannon1949}. More specifically, in each transmission round, the main and auxiliary messages at each user are splitted into several parts which include not only the secret and public parts, but also a key part and an encrypted message part. That is, at round $t,$ to send the main message $m_i^t=(m_{i, s}^t, m_{i, e}^t),$ User $i$ randomly chooses $(m_{i,k}^t, m_{i,o}^t),$ where   
\begin{enumerate}
	\item $\tilde{m}_i^t=(m_{i, s}^t, m_{i,k}^t)$ are the message parts to be kept secret (at round $t$) that include
	\begin{itemize}
		\item a secret message $m_{i, s}^t,$ where $m_{i, s}^t\in [1, e^{nR_{i,s}}],$
		\item a key $m_{i,k}^t,$ which is to be used for encryption by the other user for the next round, where $m_{i, k}^t\in [1, e^{nR_{i,k}}],$
	\end{itemize}
	\item  $\tilde{m}_{i,r}^t=(m_{i, e}^t\oplus m_{i\oplus 1,k}^{t-1}, m_{i,o}^t)$ are the message parts that could be public that include
	\begin{itemize}
		\item an encrypted message $m_{i, e}^t\oplus m_{i\oplus 1, k}^{t-1}$, where the message part $m_{i, e}^t$ is encrypted using one-time pad with the key $m_{i\oplus 1,k}^{t-1}$ from the other user from the previous round, where $m_{i, e}^t\in [1, e^{nR_{i,e}}]$ and  $m_{i\oplus 1,k}^{t-1}\in [1, e^{nR_{i\oplus 1,k}}].$ Note that to ensure the perfect secrecy \cite{Shannon1949} of  $m_{i, e}$ by one-time pad with $m_{i\oplus 1,k}^{t-1}$, it requires that $R_{i,e}\leq R_{i\oplus 1,k}.$
		\item an open message $m_{i,o}^t,$ where $m_{i, o}^t\in [1, e^{nR_{i,o}}].$
	\end{itemize}
\end{enumerate}  
Since the encoding at each round used the key received from the previous round, the coding strategy is no longer non-adaptive although the inputs still remain independent.  

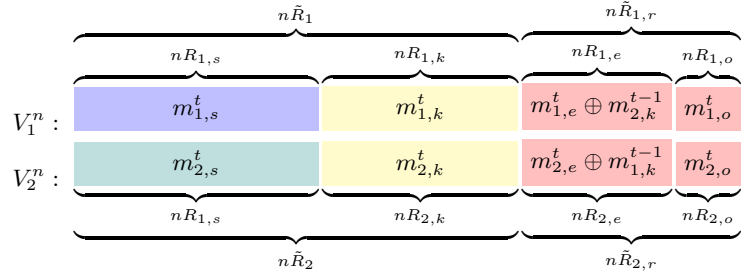
\begin{figure}[h]
	\centering	
	\begin{tabular}{rcl}		
		$V_1^n:$ &	& 
		$
		\overbrace{
		\overbrace{	
			\begin{tikzpicture}
				\node[minimum height=1.6em, minimum width=10em, anchor=base, fill=blue!25] {$m_{1,s}^t$}; 
			\end{tikzpicture}
		}^{nR_{1,s}}
		\overbrace{	
			\begin{tikzpicture}
				\node[minimum height=1.6em, minimum width=8em, anchor=base, fill=yellow!25] {$m_{1,k}^t$}; 
			\end{tikzpicture}
		}^{nR_{1,k}}
		}^{n\tilde{R}_1}
		\overbrace{
		\overbrace{
			\begin{tikzpicture}
				\node[minimum height=1.6em, minimum width=2em, anchor=base, fill=red!25] {$m_{1,e}^t\oplus m_{2,k}^{t-1}$}; 
			\end{tikzpicture}
		}^{nR_{1,e}}		
	\overbrace{
		\begin{tikzpicture}
			\node[minimum height=1.6em, minimum width=1.5em, anchor=base, fill=red!25] {$m_{1,o}^t$}; 
		\end{tikzpicture}
	}^{nR_{1,o}}
	}^{n\tilde{R}_{1,r}}
	$ \\
		$V_2^n:$ & 	& 
		$
		\underbrace{
		\underbrace{	
			\begin{tikzpicture}
				\node[minimum height=1.6em, minimum width=10em, anchor=base, fill=teal!25] {$m_{2,s}^t$}; 
			\end{tikzpicture}
		}_{nR_{1,s}}
		\underbrace{	
			\begin{tikzpicture}
				\node[minimum height=1.6em, minimum width=8em, anchor=base, fill=yellow!25] {$m_{2,k}^t$}; 
			\end{tikzpicture}
		}_{nR_{2,k}}
		}_{n\tilde{R}_{2}}
		\underbrace{
		\underbrace{
			\begin{tikzpicture}
				\node[minimum height=1.6em, minimum width=2.5em, anchor=base, fill=red!25] {$m_{2,e}^t\oplus m_{1,k}^{t-1}$}; 
			\end{tikzpicture}
		}_{nR_{2,e}}		
		\underbrace{
			\begin{tikzpicture}
				\node[minimum height=1.6em, minimum width=1.5em, anchor=base, fill=red!25] {$m_{2,o}^t$}; 
			\end{tikzpicture}
		}_{nR_{2,o}}
		}_{n\tilde{R}_{2,r}}
	$ 
	\end{tabular}
	\caption{Encoding for Adaptive Coding at Round $t$.}
	\label{fig: Message structure b}	
\end{figure}

\noindent {\it Codebook Generation:}
At transmitter $i,$ let $V_{i,1}^n,\ldots, V_{i, \sM_i \cdot \sL_i}^n$ 
be random variables independently subject to 
$P_{V_i^n}=\prod_{j=1}^{n}P_{V_{i,j}}, $ where 
$V_i^n=(V_{i,1}, \cdots, V_{i,n}),$  
$\sM_i= e^{n \tilde{R}_i}$ and 
$\sL_i= e^{n \tilde{R}_{i,r}}$ for $i=1,2.$

\noindent {\it Encoding:}
At round $t>1,$ when the legitimate user $i$ intends to send the message $m_i^t=(m_{i, s}^t, m_{i, e}^t)$, he(she) randomly chooses $m_{i,k}^t\in [1, e^{nR_{i,k}}]$ and  $m_{i, o}^t\in [1, e^{nR_{i,o}}]$ and sends ${V}^n_{i,(\tilde{m}_i^t, \tilde{m}_{i,r}^t)},$ with $\tilde{m}_i^t=(m_{i, s}^t, m_{i,k}^t)$ and $\tilde{m}_{i,r}^t=(m_{i, e}^t\oplus m_{i\oplus 1,k}^{t-1}, m_{i,o}^t)$. Here $m_{i\oplus 1,k}^{t-1}$ is the key from the other user from the previous round. 

Note that at the first round, the legitimate user $i$ sends a fixed message $m_i^1=(0, 0)$. He(she) randomly chooses $m_{i,k}^1\in [1, e^{nR_{i,k}}]$ and $m_{i,o}^1\in [1, e^{nR_{i,o}}]$ and sends ${V}^n_{i,(\tilde{m}_i^1, \tilde{m}_{i,r}^1)},$ with $\tilde{m}_i^1=(0, m_{i,k}^1)$ and $\tilde{m}_{i,r}^1=(0, m_{i,o}^1)$.

\noindent {\it Decoding:}
At the other legitimate receiver, an ML decoder will be used to decode ${V}^n_{i,(\tilde{m}_i^t, \tilde{m}_{i,r}^t)}$ by using the receiver's information $X_{i \oplus 1}^n$. Thus $m_i^t=(m_{i, s}^t, m_{i, e}^t)$ is obtained by taking $m_{i,s}^t$ from $\tilde{m}_i^t$ and taking $m_{i,e}^t$ from decoding $\tilde{m}_{i,r}^t$ with $m_{i\oplus 1,k}^{t-1}.$ Note that $m_{i\oplus 1,k}^{t-1}$ is its key message part (being kept secret from the eavesdropper) in the previous round. $m_{i,k}^{t}$ from $\tilde{m}_i^t$ will be decoded as well (and used in the next round encoding).

Note that at the first round, only the $m_{i,k}^{1}$ will be decoded. 

If we use the code $t>1$ rounds, we obtain an adaptive code of length $nt.$ We notice that in the first round, a fixed message $m_i^1=(0, 0)$ is sent. That is, at User $i,$ rate $0$ is for the first round and $R_i$ for the rest $t-1$ rounds. The overall rate at User $i$ becomes $R$ as $t$ goes to infinity:
\begin{equation}
	\lim\limits_{t\to \infty} \frac{n R_{i,s}+n(t-1)R_i}{nt} = R_i.
\end{equation} 

In this way, User $i$ is able to send the message $m_i=(m_{i,s}, m_{i,e})$ of rate $R_i,$ where
\begin{align}
	R_1 =& R_{1,s} + R_{1,e}, \Label{G8A}\\ 
	R_2 =& R_{2,s} + R_{2,e}. \Label{G9A}
\end{align}  
by employing a $(2^{n\tilde{R}_1}, 2^{n\tilde{R}_2}, n)$ non-adaptive code in $t$ rounds (together with the key exchange mechanism and one-time pad encoding in each round),  where 
\begin{align}
	\tilde{R}_i =& R_{i,s} + R_{i,k}, \Label{G11A}\\
	\tilde{R}_{i,r} =& R_{i,o} + R_{i,e}. \Label{G12A}
\end{align}
Note that $R_{i,s}$ is the rate of the secret message part; $R_{i,k}$ is the rate of the key part (which User $i$ sacrifices for the key exchange); $R_{i,o}$ is the rate of the partial messages that could be public; and $R_{i,e}$ is the rate of encrypted message part (which User $i$ benefits from the key exchange). 

In the following, for simplicity, we also denote $(M_{\star}^1, \cdots, M_{\star}^t)$ as $\mathbf{M_{\star}}[1:t]$ or simply $\mathbf{M_{\star}}.$ Moreover, we also use $\mathbf{M_{\star}}[j]$ to denote the corresponding variable in the $j$-th round, i.e., $M_{\star}^{j}.$ Note that $M_{\star}$ can be $M_{i}, M_{i,s}, M_{i,k}, M_{i,e}. M_{i,o}, Z^n$ etc.

\subsection{Exponential evaluation}
We have the following Lemma for the adaptive coding, which gives exponential evaluation on both the decoding error probability  and the information leaked to the eavesdropper. The proof is provided   in the reliability analysis in Section \ref{sec: reliability A} and strong secrecy analysis in Section \ref{sec: Secrecy A}.
\begin{lemma}\Label{L3TA}
	For a joint distribution $P_{V_1 V_2 X_1 X_2} \in {\cal Q}$,
	 positive numbers $(R_{1,s}, R_{1,k}$, $R_{1,e}, R_{1,o}, R_{2,s},R_{2,k}, R_{2,e}, R_{2,o})$, there exists a sequence of $(e^{ntR_1}, e^{ntR_2}, nt)$ adaptive codes $\Co_n^t$ with $(R_1,R_2)=(R_{1,s}+R_{1,e}, R_{2,s}+R_{2,e})$ that is constructed based on $t$ rounds of $(e^{n\tilde{R}_1}, e^{n\tilde{R}_2}, n)$ non-adaptive codes $\Co_n$ with $(\tilde{R}_1,\tilde{R}_2)=(R_{1,s}+R_{1,k}, R_{2,s}+R_{2,k})$ and $(\tilde{R}_{1,r},\tilde{R}_{2,r})=(R_{1,e}+R_{1,o}, R_{2,e}+R_{2,o})$ such that
	\begin{align} 
&		P_{e}^n(\Co_n^t) \leq \ 2t\left[e^{ns\left(  \tilde{R}_2+\tilde{R}_{2,r}-  I_{\frac{1}{1+s}}^{\downarrow} (Y_1 ;  V_{2}| X_{1})\right)}+e^{ns\left(  \tilde{R}_1+\tilde{R}_{1,r}-  I_{\frac{1}{1+s}}^{\downarrow} (Y_2 ;  V_{1}| X_{2})\right)}\right]
		, \Label{eq:ReliabilityDA} \\
&		I(\mathbf{M_{1}},\mathbf{M_{2}}; \mathbf{Z^{n}}|\Co_n^t)	\leq	2t\left[
		\sum_{\mathcal{S}\neq \emptyset, \mathcal{S}\subset  \{1,2\}}
		e^{n s\left( I^{\uparrow}_{1+s} ( Z ; V_{\mathcal{S}})-\sum\limits_{j \in \mathcal{S}}\tilde{R}_{j,r}\right) }\right]. \Label{eq:SSecDA}
	\end{align}
	Also, there also exists a sequence of $(e^{ntR_1}, e^{ntR_2}, nt)$ adaptive codes $\Co_n^t$ that is constructed based on $t$ rounds of $(e^{n\tilde{R}_1}, e^{n\tilde{R}_2}, nt)$ non-adaptive codes $\Co_n$ such that
	\begin{align} 
		&P_{e}^n(\Co_n^t) \notag \\
		\leq & 3 t\left[e^{ns\left(  \tilde{R}_2+\tilde{R}_{2,r}-  I_{\frac{1}{1+s}}^{\downarrow} (Y_1 ;  V_{2}| X_{1})\right)}+e^{ns\left(  \tilde{R}_1+\tilde{R}_{1,r}-  I_{\frac{1}{1+s}}^{\downarrow} (Y_2 ;  V_{1}| X_{2})\right)}\right]
		, \Label{eq:Reliability2DA} \\
		&I(\mathbf{M_{1}}; \mathbf{Z^{n}}|\Co_n^t)	\notag \\
		\leq &3t \Big[e^{n s\left( I^{\uparrow}_{1+s} ( Z ; V_1,V_2)-(\tilde{R}_{1,r}+\tilde{R}_{2,r}+R_{2,s})\right) }
		+e^{n s\left( I^{\uparrow}_{1+s} ( Z ; V_1)- \tilde{R}_{1,r}\right) }\notag \\
		&+e^{n s\left( I^{\uparrow}_{1+s} ( Z ; V_2)-(\tilde{R}_{2,r}+R_{2,s})\right) }\Big]
		, \Label{eq:SSec2DA} \\
		&I(\mathbf{M_{2}}; \mathbf{Z^{n}}|\Co_n^t)	\notag \\
		\leq &3t\Big[
		e^{n s\left( I^{\uparrow}_{1+s} ( Z ; V_1,V_2)-(\tilde{R}_{1,r}+\tilde{R}_{2,r}+R_{1,s})\right) }
		+e^{n s\left( I^{\uparrow}_{1+s} ( Z ; V_2)- \tilde{R}_{2,r}\right) }\notag\\
		&+e^{n s\left( I^{\uparrow}_{1+s} ( Z ; V_1)-(\tilde{R}_{1,r}+R_{1,s})\right) }\Big]
		. \Label{eq:SSec3DA}
	\end{align}
\end{lemma}

\subsection{Reliability Analysis of Adaptive Coding}\label{sec: reliability A}
Recall the analysis in \cite{HC2023}. By applying Lemma \ref{L3} to the $n$-fold asymptotic case, the average of the decoding error probability $p_{N, e}$ of using a $(2^{n\tilde{R}_1}, 2^{n\tilde{R}_2}, n)$ non-adaptive code can be upper bounded by $$e^{ns\left(\tilde{R}_2+\tilde{R}_{2,r}-  I_{\frac{1}{1+s}}^{\downarrow} (Y_1 ;  V_{2}| X_{1})\right)}+e^{ns\left(  \tilde{R}_1+\tilde{R}_{1,r}-  I_{\frac{1}{1+s}}^{\downarrow} (Y_2 ;  V_{1}| X_{2})\right)}.$$
Note that the considered adaptive coding is corresponding to $t$ rounds of non-adaptive coding. An error occurring in any round (may or may not influence the correct decoding in the following round) can be considered as an error of the adaptive coding.
Then, we have the following for the average of the decoding probability of using an adaptive code $p_{A, e}:$   
\begin{align*}
	p_{A, e} &\leq 1- (1-p_{N, e})^t \\
	  		& \stackrel{(a)}{\leq} t\cdot p_{N, e}\\
	  		& \leq t\left[e^{ns\left(  \tilde{R}_2+\tilde{R}_{2,r}-  I_{\frac{1}{1+s}}^{\downarrow} (Y_1 ;  V_{2}| X_{1})\right)}+e^{ns\left(  \tilde{R}_1+\tilde{R}_{1,r}-  I_{\frac{1}{1+s}}^{\downarrow} (Y_2 ;  V_{1}| X_{2})\right)}\right],
\end{align*}
where $(a)$ is due to the fact that $(1-x)^t \geq 1-tx$ for $x\leq 1.$

So for $t$ such that $\lim\limits_{n\to \infty}\frac{\ln t}{n}\to 0$, when  
\begin{align}
	\tilde{R}_2+\tilde{R}_{2,r} &< I(Y_1;V_2|X_1),\Label{G5A}\\
	\tilde{R}_1+\tilde{R}_{1,r} &< I(Y_2;V_1|X_2), \Label{G4A}
\end{align}
the decoding error probability $p_{A, e}$ goes to zero as $n\to \infty$.

\subsection{Strong Secrecy Analysis of Adaptive Coding}\label{sec: Secrecy A}
Recall the fact that the encoding of the adaptive code at each round uses the key received from the previous round to encrypt the partial message in the one-time pad manner. To ensure the perfect secrecy \cite{Shannon1949}, it is required that 
\begin{align}
	R_{1,e} \leq & R_{2,k}, \Label{G6A}\\
	R_{2,e} \leq & R_{1,k}. \Label{G7A}
\end{align}
We notice that the inputs at each legitimate user remain independent from the previous rounds due to the one-time pad employed in the encoding. 

\begin{figure}[htbp]
	\centering
	\includegraphics[width=.75\textwidth]{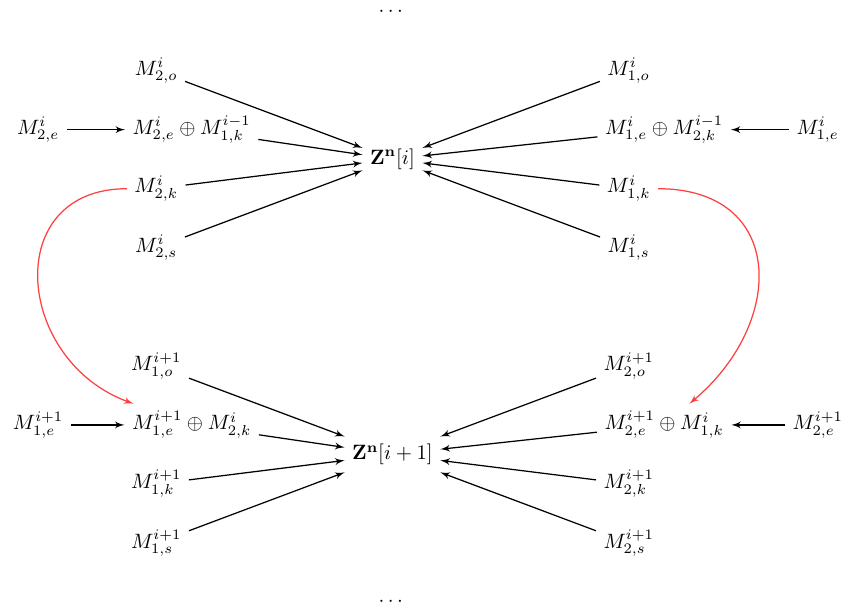}
	\caption{Dependency Graph of Adaptive Coding.} \Label{fig: Dependency}
\end{figure}

\subsubsection{Joint Secrecy}

For the strong joint secrecy of  $(M_{1}^1, \cdots, M_{1}^t, M_{2}^1, \cdots, M_{2}^t),$ i.e., $(\mathbf{M_{1}},\mathbf{M_{2}}),$ we consider 
\allowdisplaybreaks
\begin{align*}
	&I(\mathbf{M_{1}},\mathbf{M_{2}}; \mathbf{Z^{n}}|\Co_n^t) \\
	=&I(\mathbf{M_{1,s}}, \mathbf{M_{1,e}}, \mathbf{M_{2,s}}, \mathbf{M_{2,e}}; \mathbf{Z^{n}}|\Co_n^t) \\
	= & H(\mathbf{Z^{n}}|\Co_n^t)-H(\mathbf{Z^{n}}|\mathbf{M_{1,s}}, \mathbf{M_{1,e}}, \mathbf{M_{2,s}}, \mathbf{M_{2,e}}, \Co_n^t)\\
	= & \sum_{i=t}^{1} H(\mathbf{Z^{n}}[i]|\mathbf{Z^{n}}[i+1:t], \Co_n^t)\notag\\
	&- \sum_{i=t}^{1} H(\mathbf{Z^{n}}[i]|\mathbf{M_{1,s}}, \mathbf{M_{1,e}}, \mathbf{M_{2,s}}, \mathbf{M_{2,e}},\mathbf{Z^{n}}[i+1:t], \Co_n^t)\\
	\stackrel{(a)}{\leq} & \sum_{i=t}^{1} H(\mathbf{Z^{n}}[i]|\Co_n^t)\notag\\
	&- \sum_{i=t}^{1} H(\mathbf{Z^{n}}[i]|\mathbf{M_{1,s}}, \mathbf{M_{1,e}}, \mathbf{M_{2,s}}, \mathbf{M_{2,e}}, \mathbf{M_{1,k}}[i], \mathbf{M_{2,k}}[i], \mathbf{Z^{n}}[i+1:t], \Co_n^t)\\
	\stackrel{(b)}{=}& \sum_{i=t}^{1} H(\mathbf{Z^{n}}[i]|\Co_n^t)\notag \\
	&- \sum_{i=t}^{1} H(\mathbf{Z^{n}}[i]|\mathbf{M_{1,s}}[1:i], \mathbf{M_{1,e}}[1:i], \mathbf{M_{2,s}}[1:i], \mathbf{M_{2,e}}[1:i], \mathbf{M_{1,k}}[i], \mathbf{M_{2,k}}[i], \Co_n^t)\\
	\stackrel{(c)}{=}& \sum_{i=t}^{1} H(\mathbf{Z^{n}}[i]|\Co_n)- \sum_{i=t}^{1} H(\mathbf{Z^{n}}[i]|\mathbf{M_{1,s}}[i], \mathbf{M_{2,s}}[i], \mathbf{M_{1,k}}[i], \mathbf{M_{2,k}}[i], \Co_n)\\
	= & \sum_{i=t}^{1} I(\mathbf{M_{1,s}}[i], \mathbf{M_{2,s}}[i], \mathbf{M_{1,k}}[i], \mathbf{M_{2,k}}[i]; \mathbf{Z^{n}}[i]|\Co_n)\\
	= & \sum_{i=1}^{t} I(M_{1,s}^{i}, M_{2,s}^{i}, M_{1,k}^{i}, M_{2,k}^{i}; \mathbf{Z^{n}}[i]|\Co_n)\\
	=& \sum_{i=1}^{t} I(\tilde{M}_1^{i}, \tilde{M}_2^{i}; \mathbf{Z^{n}}[i]|\Co_n),
\end{align*}
where $(a)$ is due to the fact that conditioning reduces entropy; $(b)$ is due to the fact that given $(\mathbf{M_{1,k}}[i], \mathbf{M_{2,k}}[i]),$ it can be seen from the dependency graph in Fig. \ref{fig: Dependency} that $(\mathbf{Z^{n}}[i], \mathbf{M_{1,s}}[1:i], \mathbf{M_{1,e}}[1:i], \mathbf{M_{2,s}}[1:i], \mathbf{M_{2,e}}[1:i])$ is independent of $(\mathbf{Z^{n}}[i+1:t], \mathbf{M_{1,s}}[i+1:t], \mathbf{M_{1,e}}[i+1:t], \mathbf{M_{2,s}}[i+1:t], \mathbf{M_{2,e}}[i+1:t]);$ $(c)$ is due to the fact that $\mathbf{Z^{n}}[i]$ in the $i$-th round transmission is only dependent on the $(\mathbf{M_{1,k}}[i], \mathbf{M_{2,k}}[i], \mathbf{M_{1,s}}[i], \mathbf{M_{2,s}}[i])$, but not $(\mathbf{M_{1,s}}[1:i-1], \mathbf{M_{2,s}}[1:i-1])$ and $(\mathbf{M_{1,e}}[1:i],  \mathbf{M_{2,e}}[1:i]).$ 

Recall the strong joint secrecy analysis for the $(2^{n\tilde{R}_1}, 2^{n\tilde{R}_2}, n)$ non-adaptive code in \cite{HC2023}. Applying the resolvability lemma, i.e., Lemma \ref{L2}, to the $n$-fold asymptotic case, the expectation of the information leakage $I({\tilde{M}_1^i}, {\tilde{M}_2^i}; \mathbf{Z^{n}}[i]|\Co_n)$ over $\Co_n$ can be upper bounded by
$$
\sum_{\mathcal{S}\neq \emptyset, \mathcal{S}\subset  \{1,2\}}
e^{n s\left( I^{\uparrow}_{1+s} ( Z ; V_{\mathcal{S}})-\sum\limits_{j \in \mathcal{S}}\tilde{R}_{j,r}\right)}.
$$
Then the expectation of $I(\mathbf{M_{1}},\mathbf{M_{2}}; \mathbf{Z^{n}}|\Co_n^t)$ over $\Co_n^t$ can be upper bounded by 
$$
t\left[\sum_{\mathcal{S}\neq \emptyset, \mathcal{S}\subset  \{1,2\}}
e^{n s\left( I^{\uparrow}_{1+s} ( Z ; V_{\mathcal{S}})-\sum\limits_{j \in \mathcal{S}}\tilde{R}_{j,r}\right)}\right].
$$
It goes to zero as $n\to \infty$ (and for $t$ such that $\lim\limits_{n\to \infty}\frac{\ln t}{n}\to 0$), when the following rate conditions are fulfilled:
\begin{align}
	\tilde{R}_{1,r} &> I(Z;V_1); \Label{G11Y}\\
	\tilde{R}_{2,r}&> I(Z;V_2);\Label{G22Y}\\
	\tilde{R}_{1,r}+\tilde{R}_{2,r} &> I(Z;V_1, V_2).\Label{G33Y}
\end{align}

As a summary, we obtain the following rate constraints for the proposed adaptive coding under the joint strong secrecy:
\begin{itemize}
	\item rate splitting as defined in \eqref{G8A} and \eqref{G9A};
	\item rate constraints for achieving reliability as established in  \eqref{G5A} and \eqref{G4A};
	\item rate constraints for achieving perfect secrecy in one-time pad as given in \eqref{G6A} and \eqref{G7A};
	\item rate constraints for achieving strong joint secrecy as shown in \eqref{G11Y}, \eqref{G22Y} and \eqref{G33Y}.  
\end{itemize}
Replacing $(\tilde{R}_i, \tilde{R}_{i,r})$ in \eqref{G5A}, \eqref{G4A}, \eqref{G6A}, \eqref{G7A}, \eqref{G11Y}, \eqref{G22Y} and \eqref{G33Y} by the rate splitting as defined in \eqref{G11A} and \eqref{G12A}. We obtain the following set of rate inequalities (together with the non-negativity of all the rates):
\allowdisplaybreaks
\begin{align}
	R_1 =& R_{1,s} + R_{1,e},\\ 
	R_2 =& R_{2,s} + R_{2,e},\\
	R_{2,s} + R_{2,k}+R_{2,o} + R_{2,e} <& I(Y_1;V_2|X_1), \\
	R_{1,s} + R_{1,k}+R_{1,o} + R_{1,e}  <& I(Y_2;V_1|X_2),\\
	R_{1,e} \leq & R_{2,k}, \\
	R_{2,e} \leq & R_{1,k}, \\
	R_{1,o} + R_{1,e} > &I(Z;V_1), \Label{G1YA}\\
	R_{2,o} + R_{2,e} > & I(Z;V_2),\Label{G2YA}\\
	R_{1,o} + R_{1,e} +R_{2,o} + R_{2,e} > & I(Z;V_1, V_2).\Label{G3YA}
\end{align} 
Applying Fourier-Motzkin elimination to remove $R_{1,s}, R_{1,k}, R_{1,o}, R_{1,e}, R_{2,s}, R_{2,k}, \allowbreak R_{2,o}, R_{2,e}$ (see the detailed steps in Appendix \ref{APP: FM Joint}), we obtain the rate region of $(R_1,R_2)$ defined by
\begin{align}
	\begin{split}
		R_1 &\leq I(Y_2;V_1|X_2),\\
		R_2 &\leq I(Y_1;V_2|X_1),\\
		R_1+R_2 & \leq  I(Y_2;V_1|X_2) + I(Y_1;V_2|X_1) - I(V_1, V_2;Z).
	\end{split}\Label{HH1A}
\end{align}
Thus we establish the achievability of ${\cal R}_{J,A}(P_{V_1,V_2,X_1,X_2})$ as defined in \eqref{BPW1A}.

\subsubsection{Individual Secrecy}
For the strong individual secrecy of $\mathbf{M_{1}}=(M_{1}^1, \cdots, M_{1}^t)$ and $\mathbf{M_{2}}=(M_{2}^1, \cdots, M_{2}^t)$, we first consider for $\mathbf{M_{1}}$ as follows: 
\allowdisplaybreaks
\begin{align*}
&I(\mathbf{M_{1}}; \mathbf{Z^{n}}|\Co_n^t) \notag\\
	= & I(\mathbf{M_{1,s}}, \mathbf{M_{1,e}}; \mathbf{Z^{n}}|\Co_n^t) \\
	= & H(\mathbf{Z^{n}}|\Co_n^t)-H(\mathbf{Z^{n}}|\mathbf{M_{1,s}}, \mathbf{M_{1,e}}\Co_n^t)\\
	= & \sum_{i=t}^{1} H(\mathbf{Z^{n}}[i]|\mathbf{Z^{n}}[i+1:t], \Co_n^t)- \sum_{i=t}^{1} H(\mathbf{Z^{n}}[i]|\mathbf{M_{1,s}}, \mathbf{M_{1,e}},\mathbf{Z^{n}}[i+1:t], \Co_n^t)\\
	\stackrel{(a)}{\leq} & \sum_{i=t}^{1} H(\mathbf{Z^{n}}[i]|\Co_n^t)- \sum_{i=t}^{1} H(\mathbf{Z^{n}}[i]|\mathbf{M_{1,s}}, \mathbf{M_{1,e}},\mathbf{M_{1,k}}[i], \mathbf{M_{2,k}}[i], \mathbf{Z^{n}}[i+1:t], \Co_n^t)\\
	\stackrel{(b)}{=}& \sum_{i=t}^{1} H(\mathbf{Z^{n}}[i]|\Co_n^t)- \sum_{i=t}^{1} H(\mathbf{Z^{n}}[i]|\mathbf{M_{1,s}}[1:i], \mathbf{M_{1,e}}[1:i], \mathbf{M_{1,k}}[i], \mathbf{M_{2,k}}[i], \Co_n^t)\\
	\stackrel{(c)}{=}& \sum_{i=t}^{1} H(\mathbf{Z^{n}}[i]|\Co_n)- \sum_{i=t}^{1} H(\mathbf{Z^{n}}[i]|\mathbf{M_{1,s}}[i], \mathbf{M_{1,k}}[i], \mathbf{M_{2,k}}[i], \Co_n)\\
	= & \sum_{i=t}^{1} I(\mathbf{M_{1,s}}[i], \mathbf{M_{1,k}}[i], \mathbf{M_{2,k}}[i]; \mathbf{Z^{n}}[i]|\Co_n)\\
	= & \sum_{i=1}^{t} I(M_{1,s}^{i}, M_{1,k}^{i}, M_{2,k}^{i}; \mathbf{Z^{n}}[i]|\Co_n)\\
	=& \sum_{i=1}^{t} I(\tilde{M}_1^{i}, M_{2,k}^{i}; \mathbf{Z^{n}}[i]|\Co_n),
\end{align*}
where $(a)$ is due to the fact that conditioning reduces entropy; $(b)$ is due to the fact that given $(\mathbf{M_{1,k}}[i], \mathbf{M_{2,k}}[i]),$ it can be seen from the dependency graph in Fig. \ref{fig: Dependency} that $(\mathbf{Z^{n}}[i], \mathbf{M_{1,s}}[1:i], \mathbf{M_{1,e}}[1:i])$ is independent of $(\mathbf{Z^{n}}[i+1:t], \mathbf{M_{1,s}}[i+1:t], \mathbf{M_{1,e}}[i+1:t]);$ $(c)$ is due to the fact that $\mathbf{Z^{n}}[i]$ in the $i$-th round transmission is only dependent on $(\mathbf{M_{1,k}}[i], \mathbf{M_{2,k}}[i], \mathbf{M_{1,s}}[i])$, but not on $\mathbf{M_{1,s}}[1:i-1]$ and $\mathbf{M_{1,e}}[1:i].$ 

Similar to the joint secrecy analysis, we apply Lemma \ref{L2} to the $n$-fold asymptotic case for the non-adaptive code that is used in the $i$-th round of the adaptive coding. 
We obtain that the expectation of $ I(\tilde{M}_1^{i}, M_{2,k}^{i}; \mathbf{Z^{n}}[i]|\Co_n)$ over $\Co_n$ can be upper bounded by
$$e^{n s\left( I^{\uparrow}_{1+s} ( Z ; V_1,V_2)-(\tilde{R}_{1,r}+\tilde{R}_{2,r}+R_{2,s})\right) }
+e^{n s\left( I^{\uparrow}_{1+s} ( Z ; V_1)- \tilde{R}_{1,r}\right) }
+e^{n s\left( I^{\uparrow}_{1+s} ( Z ; V_2)-(\tilde{R}_{2,r}+R_{2,s})\right)}.$$
Then the expectation of $I(\mathbf{M_{1}}; \mathbf{Z^{n}}|\Co_n^t)$ over $\Co_n^t$ can be upper bounded by 
$$
t\left[e^{n s\left( I^{\uparrow}_{1+s} ( Z ; V_1,V_2)-(\tilde{R}_{1,r}+\tilde{R}_{2,r}+R_{2,s})\right) }
+e^{n s\left( I^{\uparrow}_{1+s} ( Z ; V_1)- \tilde{R}_{1,r}\right) }
+e^{n s\left( I^{\uparrow}_{1+s} ( Z ; V_2)-(\tilde{R}_{2,r}+R_{2,s})\right) }\right].
$$
It goes to zero as $n\to \infty$ (and for $t$ such that $\lim\limits_{n\to \infty}\frac{\ln t}{n}\to 0$), when the following rate conditions are fulfilled:
\begin{align*}
	\tilde{R}_{1,r} &> I(Z;V_1); \\
	\tilde{R}_{2,r}+R_{2,s}&> I(Z;V_2);\\
	\tilde{R}_{1,r}+\tilde{R}_{2,r}+R_{2,s} &> I(Z;V_1, V_2).
\end{align*}

Since the same discussion holds for 
the mutual information $ I(\tilde{M}_2^{i}, M_{1,k}^{i}; \mathbf{Z^{n}}[i]|\Co_n)$
for strong individual secrecy of $\mathbf{M_{2}}=(M_{2}^1, \cdots, M_{2}^t)$,  we have 
\begin{align*}
	\tilde{R}_{2,r} &> I(Z;V_2);\\
	\tilde{R}_{1,r}+R_{1,s} &> I(Z;V_1);\\
	\tilde{R}_{1,r}+\tilde{R}_{2,r}+R_{1,s} &> I(Z;V_1, V_2).
\end{align*}

Together we have the following constraints for the strong individual secrecy of $\mathbf{M_{1}}=(M_{1}^1, \cdots, M_{1}^t)$ and $\mathbf{M_{2}}=(M_{2}^1, \cdots, M_{2}^t)$:
\begin{align}
	\tilde{R}_{1,r} &> I(Z;V_1);\Label{G1A}\\
	\tilde{R}_{2,r} &> I(Z;V_2);\Label{G2A}\\
	\tilde{R}_{1,r}+\tilde{R}_{2,r} +\min(R_{1,s},R_{2,s}) &> I(Z;V_1, V_2).\Label{G3A}
\end{align}

As a summary, we obtain the following rate constraints for the proposed adaptive coding under the joint individual secrecy:
\begin{itemize}
	\item rate splitting as defined in \eqref{G8A} and \eqref{G9A};
	\item rate constraints for achieving reliability as established in  \eqref{G5A} and \eqref{G4A};
	\item rate constraints for achieving perfect secrecy in one-time pad as given in \eqref{G6A} and \eqref{G7A};
	\item rate constraints for achieving strong individual secrecy as shown in \eqref{G1A}, \eqref{G2A} and \eqref{G3A}.  
\end{itemize}
Replacing $(\tilde{R}_i, \tilde{R}_{i,r})$ in \eqref{G5A}, \eqref{G4A}, \eqref{G6A}, \eqref{G7A}, \eqref{G1A}, \eqref{G2A} and \eqref{G3A} by the rate splitting as defined in \eqref{G11A} and \eqref{G12A}. We obtain the following set of rate inequalities (together with the non-negativity of all the rates):
\allowdisplaybreaks
\begin{align}
	R_1 =& R_{1,s} + R_{1,e},\\ 
	R_2 =& R_{2,s} + R_{2,e},\\
	R_{2,s} + R_{2,k}+R_{2,o} + R_{2,e} <& I(Y_1;V_2|X_1), \\
	R_{1,s} + R_{1,k}+R_{1,o} + R_{1,e}  <& I(Y_2;V_1|X_2),\\
	R_{1,e} \leq & R_{2,k}, \\
	R_{2,e} \leq & R_{1,k}, \\
	R_{1,o} + R_{1,e} > &I(Z;V_1),\\
	R_{2,o} + R_{2,e} > & I(Z;V_2),\\
	R_{1,o} + R_{1,e} +R_{2,o} + R_{2,e}+ \min(R_{1,s}, R_{2,s}) > & I(Z;V_1, V_2).
\end{align} 

Applying Fourier-Motzkin elimination to remove $R_{1,s}, R_{1,k}, R_{1,o}, R_{1,e}, R_{2,s}, R_{2,k}, \allowbreak R_{2,o}, R_{2,e}$ (see the detailed steps in Appendix \ref{APP: FM Individual}), we obtain the rate region of $(R_1, R_2)$ defined by
\begin{align}
	\begin{split}
		R_1 &\leq I(Y_2;V_1|X_2)\\
		R_2 &\leq I(Y_1;V_2|X_1)\\
		\max\{R_1, R_2\} &\leq  I(Y_2;V_1|X_2) + I(Y_1;V_2|X_1) - I(V_1, V_2;Z),\\
		R_1 + R_2 &
		\leq  \left\{
		\begin{array}{c}
			\min\left\{
			\begin{array}{l}
				I(Y_2;V_1|X_2)-I(Z;V_1),\\
				I(Y_2;V_1|X_2) + I(Y_1;V_2|X_1) - I(V_1, V_2;Z)
			\end{array}
			\right\}\\
			+ \\
			\min\left\{
			\begin{array}{l}
				I(Y_1;V_2|X_1)-I(Z;V_2),\\
				I(Y_2;V_1|X_2) + I(Y_1;V_2|X_1) - I(V_1, V_2;Z)
			\end{array}
			\right\}\\
		\end{array}
		\right\}
	\end{split}\Label{HH2A}
\end{align}
Thus we establish the achievability of ${\cal R}_{I,A}(P_{V_1,V_2,X_1,X_2})$ as defined in \eqref{BPW2A}.

\begin{remark}
	As being observed in \cite{PB2011} for the joint secrecy region, key exchange improves the individual bounds but not on the sum-rate (see ${\cal R}_{J,A}$ in \eqref{eqn: RJA} and  ${\cal R}_{J,N}$ in \eqref{eqn: RJN}). This applies to the individual secrecy regions as well (see ${\cal R}_{J,A}$ in \eqref{eqn: RIA} and  ${\cal R}_{J,N}$ in \eqref{eqn: RIN}). We also notice that our strong individual secrecy region improves the weak individual secrecy region as given in \cite[Theorem 1]{QDT2017}.
\end{remark}

\section{Exploring the ouputs correlation}\label{sec: Exploring the correlation}
The adaptive coding discussed in Section \ref{sec: Adaptive coding} introduces dependency between transmissions by the key exchange. The idea is to use the key received from the previous transmission round to encrypt the partial message to be transmitted. In this section, we consider a different adaptive method that explores the correlation among $Y_1,Y_2$, and $Z$ under fixed $X_1,X_2$.

\subsection{Virtual protocol}
As a condition that the second type of adaptive method works well, 
we introduce 
the conditionally independent condition for an MAC $P_{Y_1,Y_2,Z|X_1,X_2}$.
That is, we focus on 
the correlation among $Y_1,Y_2$, and $Z$ 
under fixed $X_1,X_2$.

\begin{definition}
When the MAC $P_{Y_1,Y_2,Z|X_1,X_2}$
satisfies the condition
\begin{align}
&P_{Y_1,Y_2,Z|X_1,X_2}(y_1,y_2,z|x_1,x_2)\notag\\
=&
P_{Y_1|X_1,X_2}(y_1|x_1,x_2)
P_{Y_2|X_1,X_2}(y_2|x_1,x_2)
P_{Z|X_1,X_2}(z|x_1,x_2),
\end{align}
we say that the MAC $P_{Y_1,Y_2,Z|X_1,X_2}$
is {\it conditionally independent}.
\end{definition}

To state our protocol for our MAC $P_{Y_1,Y_2,Z|X_1,X_2}$, 
we first state a virtual protocol,
which requires noiseless public channel between 
two users.
Our virtual protocol is a two-way wiretap channel version of
the existing methods presented in \cite{HV1,HV2,H22}.
We assume that 
${\cal X}_1$, ${\cal X}_2$, ${\cal Y}_1$, ${\cal Y}_2$
are finite fields.

We choose distributions $P_{X_1}$ on ${\cal X}_1$, $P_{X_2}$
on ${\cal X}_2$.
First, User 1 and User 2 generate random variables $X_1$ and $X_2$
independently, 
and use them as the channel input.
Then, User $i$ receives $Y_i$ for $i=1,2$.
Also, the adversary obtains the output $Z$.
Second, User $i$ chooses the random variable
$(X_i',Y_i')$ in ${\cal X}_i\times {\cal Y}_i$, and
sends $(X_i'',Y_i''):=  (X_i\oplus X_i',Y_i\oplus Y_i')$ 
to the other user via noiseless public channel for $i=1,2$.

Now, we consider that the first step to generate
$X_1,Y_1,X_2,Y_2,Z$ is the preparation stage,
and only the second step is considered as 
the two-way wiretap channel.
Notice that 
the distributions for $X_1,X_2$ are fixed, but 
we have no condition to generate $X_1',Y_1',X_2',Y_2'$.
Therefore, to summary up,
we consider the following MAC.
The variables $(X_1',Y_1')$
and $(X_2',Y_2')$ are 
the inputs of User $1$ and User $2$, respectively.
The variables $(X_1,Y_1,X_2'',Y_2'')$ are 
the output of Users $1$.
The variables $(X_2,Y_2,X_1'',Y_1'')$ are 
the output of Users $2$.
The variables $(Z,X_1'',Y_1'',X_2'',Y_2'')$ are 
the output of the adversary.
In this scenario, $X_1,Y_1$ need to be considered as 
a part of the output of User $1$
because they also reflect the inputs $(X_2',Y_2')$.

Then, we have the MAC
$$P_{(X_1,Y_1,X_2'',Y_2''),
(X_2,Y_2,X_1'',Y_1''), (Z,X_1'',Y_1'',X_2'',Y_2'')
|(X_1',Y_1'),(X_2',Y_2')},$$ which depends on 
the distributions $P_{X_1}$ and $P_{X_2}$.
Repeating the above procedure $n$ times, 
we apply the adaptive code for MAC channel
to the above MAC.
For this application, we need to choose 
distributions
$P_{V_1,X_1',Y_1'}$ 
and $P_{V_2,X_2',Y_2'}$.
We denote the capacity region given in \eqref{BPW1A} 
and \eqref{BPW2A}
by 
$R_{J,A}(P_{V_1,X_1',Y_1'},P_{V_2,X_2',Y_2'}|P_{X_1},P_{X_2})$
and 
$R_{I,A}(P_{V_1,X_1',Y_1'},P_{V_2,X_2',Y_2'}|P_{X_1},P_{X_2})$,
respectively.
A code $\Phi$ for the MAC
$P_{(X_1,Y_1,X_2'',Y_2''),
(X_2,Y_2,X_1'',Y_1''), (Z,X_1'',Y_1'',X_2'',Y_2'')
|(X_1',Y_1'),(X_2',Y_2')}$ with $n$ channel uses
is also considered as 
a code for the above virtual protocol with $n$ uses of
the MAC $P_{Y_1,Y_2,Z|X_1,X_2}$
with distributions $P_{X_1}$ and $P_{X_2}$.

In particular, when 
${\cal V}_i={\cal X}_i\times {\cal Y}_i$
and 
$P_{X_i',Y_i'|V_i}$ is given as
\begin{align}
P_{X_i',Y_i'|V_i}(x_i',y_i'|x_i'',y_i'' )=
\delta_{(x_i',y_i'),(x_i'',y_i'' )}
\end{align} 
for $i=1,2$,
the regions
$R_{J,A}(P_{V_1,X_1',Y_1'},P_{V_2,X_2',Y_2'}|P_{X_1},P_{X_2})$
and 
$R_{I,A}(P_{V_1,X_1',Y_1'}$, $P_{V_2,X_2',Y_2'}|P_{X_1},P_{X_2})$,
are determined by the distributions 
$P_{X_1',Y_1'},P_{X_2',Y_2'},P_{X_1},P_{X_2}$.
Then, we denote them
by 
$R_{J,A}(P_{X_1',Y_1'},P_{X_2',Y_2'}|P_{X_1},P_{X_2})$
and 
$R_{I,A}(P_{X_1',Y_1'},P_{X_2',Y_2'}|$ $P_{X_1},P_{X_2})$,
respectively. Note that here the delta function $\delta_{x,x'}$ is defined as 
\begin{align}
	\delta_{x,x'}=
	\left\{
	\begin{array}{cc}
		1 & \hbox{when } x=x' \\
		0 & \hbox{when } x \neq x' .
	\end{array}
	\right.
\end{align}

\begin{remark}
When we have only a wiretap channel with 
a single legitimate sender and a single legitimate receiver,
User 1 generates $X$ and 
User 2 receives $Y$ in the first step. 
In this case, User 2 generates
another variable $Y'$ and sends $Y''=Y\oplus Y'$ to User 1.
This two-step method was discussed in \cite{HV1,HV2,H22}.
To cover the two-way wiretap version of 
the case \cite{HV1,HV2}, 
the MAC $P_{Y_1,Y_2,Z|X_1,X_2}$ needs to satisfy the conditionally independent condition.

When the first step has been already done, 
the variables $X,Y,Z$ are already given.
When User 2 wants to send secure message to User 1 under this setting,
we can consider the situation
that User 2 generates
another variable $Y'$ and sends $Y''=Y\oplus Y'$ to User 1
in the above way.
This idea with a given joint distribution of variables $X,Y,Z$
was studied in \cite[Section VI]{Ha11}, 
\cite[Section IX]{H13}.
\end{remark}

\begin{theorem}\label{Lem3}
Assume that the MAC $P_{Y_1,Y_2,Z|X_1,X_2}$
is conditionally independent.
Assume that 
${\cal V}_1={\cal Y}_1$ and
${\cal V}_2=\{0\}$.
We also assume that
\begin{align}
P_{X_1',Y_1'|V_1}(x_1,y_1|v_1)=
\frac{1}{|{\cal Y}_1|}\delta_{x_1,0} \delta_{y_1,v_1} 
\end{align}
and the distribution $P_{X_2',Y_2'|V_2=0}$
is the uniform distribution on 
${\cal X}_2\times {\cal Y}_2$.
Then, we have
\begin{align}
&I(V_1 ; X_2,Y_2,X_1'',Y_1''| X_2',Y_2' )
- I(V_1,V_2; Z,X_1'',Y_1'',X_2'',Y_2'' ) \notag\\
&=
I(Y_1 ; X_2|X_1 ) - I(Y_1; Z|X_1)>0\label{NNK1}\\
&I(V_2 ; X_1,Y_1,X_2'',Y_2''| X_1',Y_1' )=0.\label{NNK2}
\end{align}
The same result holds by exchanging the roles of User 1 and User 2.
\end{theorem}

When we use $n$ times the channel, 
we divide $n$ channel uses into two groups equally and 
the first group is used the above way and 
the second group is used by exchanging the roles of User 1 and User 2.
Considering the above strategy, we find that
$R_{J,A}(P_{V_1,X_1',Y_1'},P_{V_2,X_2',Y_2'}|P_{X_1},P_{X_2})$
and 
$R_{I,A}(P_{V_1,X_1',Y_1'},P_{V_2,X_2',Y_2'}|P_{X_1},P_{X_2})$
have strictly positive rate region at the first component.

When we replace the role of User 1 and User 2,
$R_{J,A}(P_{V_1,X_1',Y_1'},P_{V_2,X_2',Y_2'}|P_{X_1},P_{X_2})$
and 
$R_{I,A}(P_{V_1,X_1',Y_1'},P_{V_2,X_2',Y_2'}|P_{X_1},P_{X_2})$
have strictly positive rate region at the second component.
Thus, for 
a conditionally independent MAC $P_{Y_1,Y_2,Z|X_1,X_2}$,
the regions
\begin{align}
cl. \bigcup_{ P_{V_1,X_1',Y_1'} \in {\cal Q}_1,P_{V_2,X_2',Y_2'}
\in {\cal Q}_2}
R_{J,A}(P_{V_1,X_1',Y_1'},P_{V_2,X_2',Y_2'}|P_{X_1},P_{X_2}) \\
cl. \bigcup_{ P_{V_1,X_1',Y_1'} \in {\cal Q}_1,P_{V_2,X_2',Y_2'}
\in {\cal Q}_2}
R_{I,A}(P_{V_1,X_1',Y_1'},P_{V_2,X_2',Y_2'}|P_{X_1},P_{X_2}) 
\end{align}
have strictly positive rate region for both components,
where ${\cal Q}_i$ is the set of any joint distribution
of $V_i,X_i',Y_i'$.

\begin{proof}
Since $V_2$ takes only one value $0$,
we have \eqref{NNK2}.
For \eqref{NNK1}, we have
\allowdisplaybreaks
\begin{align}
&I(V_1 ; X_2,Y_2,X_1'',Y_1''| X_2',Y_2' )
- I(V_1,V_2; Z,X_1'',Y_1'',X_2'',Y_2'' ) \notag\\
\stackrel{(a)}{=}
&I(Y_1' ; X_2,Y_2,X_1'',Y_1''| X_2',Y_2' )
- I(Y_1'; Z,X_1'',Y_1'',X_2'',Y_2'' ) \notag\\
\stackrel{(b)}{=}
&I(Y_1' ; X_2,Y_2,X_1,Y_1''| X_2',Y_2' )
- I(Y_1'; Z,X_1,Y_1'' ) \notag\\
=&
I(Y_1' ; Y_1''| X_2',Y_2' )
+I(Y_1' ; X_2,Y_2,X_1| Y_1'',X_2',Y_2' )
- I(Y_1' ; Y_1'' )
- I(Y_1'; Z,X_1|Y_1'' ) \notag\\
\stackrel{(c)}{=}
&
I(Y_1' ; Y_1'' )
+I(Y_1' ; X_2,Y_2,X_1| Y_1'',X_2',Y_2' )
- I(Y_1' ; Y_1'' )
- I(Y_1'; Z,X_1|Y_1'' ) \notag\\
\stackrel{(d)}{=}&
I(Y_1 ; X_2,Y_2,X_1| Y_1'',X_2',Y_2' )
- I(Y_1; Z,X_1|Y_1'' ) \notag\\
\stackrel{(e)}{=}&
I(Y_1 ; X_2,Y_2,X_1 )
- I(Y_1; Z,X_1) \notag\\
=&
I(Y_1 ; X_1 )
+I(Y_1 ; X_2,Y_2|X_1 )
-I(Y_1 ; X_1 )
- I(Y_1; Z|X_1) \notag\\
=&
I(Y_1 ; X_2,Y_2|X_1 )
- I(Y_1; Z|X_1) \notag\\
\stackrel{(f)}{=}&
I(Y_1 ; X_2|X_1 ) - I(Y_1; Z|X_1),
\end{align}
where
$(a)$ follows from $V_1=Y_1'$ and $V_2=0$;
$(c)$ follows from 
the independence of $X_2',Y_2'$ from $Y_1'$ and $Y_1$;
$(d)$ follows from the relation $Y_1''=Y_1+Y_1'$;
$(f)$ follows from 
Markovian chain $ Y_1-(X_1,X_2)-Y_2$.

Note that $(b)$ can be shown as follows.
Since $X_1'=0$, we have the relation $X_1''=X_1$.
Since $X_2',Y_2'$ are subject to the uniform distribution,
under the condition that $Y_1', Z,X_1'',Y_1'',X_2,Y_2$
are fixed to $y_1', z,x_1'',y_1'',x_2,y_2$,
$X_2'',Y_2''$ are subject to the uniform distribution,
under the condition that $Y_1', Z,X_1'',Y_1''$
are fixed to $y_1', z,x_1'',y_1'',x_2,y_2$.
Hence, $X_2'',Y_2''$ are independent of 
$Y_1', Z,X_1'',Y_1''$.
Thus, we obtain the step $(b)$.

Moreover, $(e)$ can be shown as follows.
Since $X_2',Y_2',Y_1'$ are subject to the uniform distribution,
under the condition that $X_1 ,Y_1, X_2,Y_2,Z$
are fixed to $x_1 ,y_1, x_2,y_2,z$,
$X_2',Y_2',Y_1''$ are subject to the uniform distribution,
under the condition that $X_1 ,Y_1, X_2,Y_2,Z$
are fixed to $x_1 ,y_1, x_2,y_2,z$.
Hence, $X_2',Y_2',Y_1''$ are independent of
$X_1 ,Y_1, X_2,Y_2,Z$.
Thus, we obtain the step $(e)$.

Further, 
since the MAC $P_{Y_1,Y_2,Z|X_1,X_2}$
is conditionally independent,
we have Markovian chain $ Y_1-(X_1,X_2)-Z$, which implies 
the inequality
$I(Y_1 ; X_2|X_1 ) > I(Y_1; Z|X_1).$
\end{proof}

\subsection{Conversion from virtual protocol}
Next, we discuss how to convert the above 
virtual protocol 
to 
a two-way channel protocol over
the given 
MAC $P_{Y_1,Y_2,Z|X_1,X_2}$.
We consider the above virtual protocol
with $n_1$ uses of the MAC $P_{Y_1,Y_2,Z|X_1,X_2}$
with the distributions $P_{X_1}$, $P_{X_2},$
whose code is denoted as $\Phi_1=(\phi_{1,1},\phi_{2,1},
\psi_{1,1},\psi_{2,1})$,
where $ \phi_{i,1}$ and $ \psi_{i,1}$ are 
User $i$'s encoder and decoder.
Under the code $\Phi_1$,
we denote 
the message from User 1 to User 2 by $M_1$,
and 
the message from User 2 to User 1 by $M_2$.
After the first stage,
User $i$ has random variables
$(X_{i,1},Y_{i,1}), \ldots, (X_{i,n_1},Y_{i,n_1})$.

In this virtual protocol with the code $\Phi_1$,
User $i$ sends 
the variables 
$(X_{i,1}'',Y_{i,1}'')$, $\ldots, (X_{i,n_1}'',Y_{i,n_1}'')$
to the other user via public channel for $i=1,2$.
That is, 
when the encoder of the code $\Phi_1$, $\phi_{i,1}$, converts
$M_i$ to the variables
$(X_{i,1}',Y_{i,1}')$, $\ldots, (X_{i,n_1}',Y_{i,n_1}')$, and 
the variables $(X_{i,1}'',Y_{i,1}'')$, $\ldots$, $(X_{i,n_1}'',Y_{i,n_1}'')$
are given as
$(X_{i,1}\oplus X_{i,1}',Y_{i,1}\oplus Y_{i,1}'), 
\ldots, (X_{i,n_1}\oplus X_{i,n_1}',Y_{i,n_1}\oplus Y_{i,n_1}')$, User $i\oplus 1$ recovers $M_{i\oplus 1}$ by applying the decoder $ \psi_{i,1}$
to 
$(X_{i\oplus 1,1},Y_{i\oplus 1,1}), \ldots, (X_{i\oplus 1,n_1},Y_{i\oplus 1,n_1})$
and 
the variables $(X_{i,1}'',Y_{i,1}'')$, $\ldots$, $(X_{i,n_1}'',Y_{i,n_1}'')$.

We denote the required bit lengths 
of the public noiseless channel 
for sending 
$(X_{1,1}'',Y_{1,1}'')$, $\ldots$, $(X_{1,n_1}'',Y_{1,n_1}'')$
and
$(X_{2,1}'',Y_{2,1}'')$, $\ldots$, $(X_{2,n_1}'',Y_{2,n_1}'')$
by $n_{1,2}$ and $n_{2,1}$, respectively.

Assume that we use the given 
MAC $P_{Y_1,Y_2,Z|X_1,X_2}$ at
$n_1+n_2$ times,
where the length $n_2$ will be chosen later.
That is, this protocol is composed of two rounds, where
the first round uses the MAC $P_{Y_1,Y_2,Z|X_1,X_2}$ at $n_1$ times,
and
the second uses the MAC $P_{Y_1,Y_2,Z|X_1,X_2}$ at $n_2$ times.
In the first round,
User $i$ generates $X_i$ according to $P_{X_i}$
independently, and receives $Y_i$ for $i=1,2$.
Two users repeat this procedure $n_1$ times.

For the second round,
two users prepare a code $\Phi_2
=(\phi_{1,2},\phi_{2,2},
\psi_{1,2},\psi_{2,2})$
for $n_2$ uses of 
the MAC $P_{Y_1,Y_2|X_1,X_2}$ 
for almost error-free communication between them.
Also, we choose the integer $n_2$ such that
User $1$ sends $n_{1,2}$ bits to User $2$ 
and
User $2$ sends $n_{2,1}$ bits to User $1$ 
under the code $\Phi_2$. 
In the second round,
$(X_{i,1}'',Y_{i,1}''), \ldots, (X_{i,n_1}'',Y_{i,n_1}'')$
is treated as the message from User $i$ to the other user
for the code $\Phi_2$. 
That is,
$X_{i,n_1+1 },\ldots, X_{i,n_1+n_2}$
are chosen as 
$\phi_{i,2}\big((X_{i,1}'',Y_{i,1}'')$, $\ldots, (X_{i,n_1}'',Y_{i,n_1}'')\big)$.
Hence, we have
\begin{align}
(X_{i,n_1+1 },\ldots, X_{i,n_1+n_2})
=\phi_{i,2}\big( 
((X_{i,1},Y_{i,1}), \ldots, (X_{i,n_1},Y_{i,n_1}))
\oplus \phi_{i,1}(M_i)
\big).
\end{align}

The decoding process for $M_i$ is given as follows.
User $i\oplus 1$ recovers
$((X_{i,1}'',Y_{i,1}'')$, $\ldots, (X_{i,n_1}'',Y_{i,n_1}''))$ by
\begin{align}
&((\tilde{X}_{i,1}'',\tilde{Y}_{i,1}''), \ldots, 
(\tilde{X}_{i,n_1}'',\tilde{Y}_{i,n_1}''))\notag\\
=&
\psi_{i,2}\big(
(X_{i \oplus 1,n_1+1},Y_{i \oplus 1,n_1+1}),
\ldots,
(X_{i \oplus 1,n_1+n_2},Y_{i \oplus 1,n_1+n_2})\big).
\end{align}
Then, using the above decoding, 
User $i\oplus 1$ recovers $M_i$ as
\begin{align}
&\psi_{i,1}\big(
(X_{i \oplus 1,1},Y_{i \oplus 1,1}),
\ldots,
(X_{i \oplus 1,n_1},Y_{i \oplus 1,n_1}),
(\tilde{X}_{i,1}'',\tilde{Y}_{i ,1}''),
\ldots, ( \tilde{X}_{i,n_1}'',\tilde{Y}_{i ,n_1}'')\big) \notag\\
=&\psi_{i,1}\big(
(X_{i \oplus 1,1},Y_{i \oplus 1,1}),
\ldots,
(X_{i \oplus 1,n_1},Y_{i \oplus 1,n_1}),\notag \\
&\psi_{i,2}(
(X_{i \oplus 1,n_1+1},Y_{i \oplus 1,n_1+1}),
\ldots,
(X_{i \oplus 1,n_1+n_2},Y_{i \oplus 1,n_1+n_2}))\big).
\end{align}

Since the calculation of the ratio $n_1/n_2$ is not easy,
it is not easy to calculate the achievable rate region 
by this method.
However, since the ratio $n_1/n_2$ is not zero even with the asymptotic limit,
when the MAC $P_{Y_1,Y_2,Z|X_1,X_2}$
is conditionally independent,
Theorem \ref{Lem3} guarantees that 
the above method realizes a positive rate region.
In contrast, 
the achievable rate region in Section \ref{2B}
is not necessarily positive 
even under the conditional independence condition.

\subsection{Application to additive two-way wire-tap channel}
In this section, we apply the above method to additive two-way wire-tap channel.
For this specific channel, we can calculate the exact values 
 $I(Y_1 ; X_2|X_1 ) - I(Y_1;Z|X_1)$ and $
 I(Y_2 ; X_1|X_2 ) - I(Y_2;Z|X_2)$, which appear in Theorem \ref{Lem3}.

In additive two-way wiretap channel, 
$X_1,X_2,Y_1,Y_2,Z$ are variables in the finite field $\FF_q$.
Then, we have
\begin{align}
Y_1&= a_1 X_1+b_1 X_2+N_1 \Label{AD1}\\
Y_2&= a_2 X_1+ b_2 X_2+N_2 \Label{AD2}\\
Z&=  a_3 X_1+b_3 X_2+N_3 ,\Label{AD3}
\end{align}
where $a_1,b_1,a_2,b_2,a_3,b_3$ are non-zero elements of $\FF_q$.
$N_1$, $N_2$, $N_3$ are the independent 
random variables in $\FF_q,$ which are independent of $X_1, X_2$.

When $X_1$ and $X_2$ are independent random variables and both have a uniform distribution over $\FF_q$, we have
\begin{align}
I(Y_2;X_1|X_2)&=\log q -H(N_2) \\
I(Y_1;X_2|X_1)&=\log q -H(N_1) \\
I(Z;X_1)&=I(Z;X_2)=0 \\
I(X_1, X_2;Z)&=\log q -H(N_3),
\end{align}
where $H(X)$ expresses the Shannon entropy of a random variable $X$
defined as $-\sum\limits_{x}P_X(x) \log P_X(x)$.
Since 
$ Y_1= b_1b_3^{-1}Z +(a_1 -b_1b_3^{-1} a_3)X_1+N_1-b_1b_3^{-1} N_3 $
and $ Y_2= a_2 a_3^{-1} Z +(b_2 -a_2 a_3^{-1} a_3)X_2+N_2-a_2 a_3^{-1} N_3 $,
we have
\begin{align}
I(Y_1;Z|X_1)&=\log q -H(N_1-b_1b_3^{-1} N_3) \\
I(Y_2;Z|X_2)&=\log q -H(N_2-a_2 a_3^{-1} N_3) .
\end{align}
Therefore, we have
\begin{align}
I(Y_1 ; X_2|X_1 ) - I(Y_1;Z|X_1)&=H(N_1-b_1b_3^{-1} N_3) -H(N_1)
\\
I(Y_2 ; X_1|X_2 ) - I(Y_2;Z|X_2)&=H(N_2-a_2 a_3^{-1} N_3) -H(N_2).
\end{align}
Since $N_3$ is independent of $N_1$ and $N_2$,
the above quantities are strictly positive, which was shown in Theorem \ref{Lem3}
in a general setting.

\section{Conclusion}\label{Sec: Conclusion}
In this paper, we studied the adaptive coding for the two-way wiretap channel. First, we considered the adaptive coding that was proposed in \cite{GKYG2013}. This adaptive coding is well structured by running a non-adaptive coding in several rounds together with a key-exchange mechanism embedded in each transmission round. The dependency is introduced by the key exchange in the manner that the legitimate user adapts the key that is received in the previous round in encrypting part of the message to be transmitted in the current round. We are able to characterize the decoding error probability and information leakage in terms of the conditional R\'enyi mutual information based on the results obtained for the underneath non-adaptive coding in \cite{HC2023}. 

Besides, we also discussed a generic adaptive coding that explores the correlation among the outputs at the receivers and takes advantage of a noiseless public channel. We show that for the two-way wiretap channel that fulfills the conditionally independent condition, positive transmission rates can be always guaranteed even under the joint secrecy constraint when allowing such an adaptive coding. This is different from the adaptive strategy that was discussed in \cite{PB2011} that assumed the existence of a discrete memoryless source (so as to take the advantage of the key generated from the source with rate-limited public communication). Both generic adaptive approaches discussed in this paper and in  \cite{PB2011} leave it as an open problem to give a closed-form characterization of the achievable secrecy rate regions.


\appendix

\section{Fourier-Motzkin elimination to derive the joint secrecy region}\Label{APP: FM Joint}
To derive the joint secrecy region, recall that we have the following rate constraints:
\allowdisplaybreaks
\begin{align}
	R_1 =& R_{1,s} + R_{1,e}, \Label{FM_Joint_8}\\ 
	R_2 =& R_{2,s} + R_{2,e},\Label{FM_Joint_9}\\
	R_{2,s} + R_{2,k}+R_{2,o} + R_{2,e} <& I(Y_1;V_2|X_1), \Label{FM_Joint_1}\\
	R_{1,s} + R_{1,k}+R_{1,o} + R_{1,e}  <& I(Y_2;V_1|X_2),\Label{FM_Joint_2}\\
	R_{1,e} \leq & R_{2,k}, \Label{FM_Joint_6}\\
	R_{2,e} \leq & R_{1,k}, \Label{FM_Joint_7}\\
	R_{1,o} + R_{1,e} > &I(Z;V_1), \Label{FM_Joint_3}\\
	R_{2,o} + R_{2,e} > & I(Z;V_2),\Label{FM_Joint_4}\\
	R_{1,o} + R_{1,e} +R_{2,o} + R_{2,e} > & I(Z;V_1, V_2).\Label{FM_Joint_5}
\end{align}
First consider \eqref{FM_Joint_2}, \eqref{FM_Joint_3} and \eqref{FM_Joint_5} to remove $R_{1,o}.$ We obtain
\begin{align}
	R_{1,s} + R_{1,k} <& I(Y_2;V_1|X_2)-I(Z;V_1),\Label{FM_Joint_10}\\
	R_{2,o} + R_{2,e} -	R_{1,s} - R_{1,k}  >&  I(Z;V_1, V_2)-I(Y_2;V_1|X_2),\Label{FM_Joint_11}\\
	R_{1,s} + R_{1,k}+ R_{1,e}  <& I(Y_2;V_1|X_2).\Label{FM_Joint_12}
\end{align}
Consider \eqref{FM_Joint_1}, \eqref{FM_Joint_4} and \eqref{FM_Joint_11} to remove $R_{2,o}.$ We obtain
\begin{align}
	R_{2,s} + R_{2,k} <& I(Y_1;V_2|X_1)-I(Z;V_2),\Label{FM_Joint_13}\\
	R_{2,s} + R_{2,k} + R_{1,s} + R_{1,k}  <& I(Y_1;V_2|X_1)+ I(Y_2;V_1|X_2)-I(Z;V_1, V_2),\Label{FM_Joint_14}\\
	R_{2,s} + R_{2,k}+R_{2,e} <& I(Y_1;V_2|X_1). \Label{FM_Joint_15}
\end{align} 
Consider \eqref{FM_Joint_7}, \eqref{FM_Joint_10},  \eqref{FM_Joint_12} and \eqref{FM_Joint_14} to remove $R_{1,k}.$ We obtain
\begin{align}
	R_{1,s} + R_{2,e} <& I(Y_2;V_1|X_2)-I(Z;V_1),\Label{FM_Joint_16}\\
	R_{1,s} + R_{1,e}+ R_{2,e}  <& I(Y_2;V_1|X_2),\Label{FM_Joint_17}\\
	R_{2,s} + R_{2,k} + R_{1,s} + R_{2,e}  <& I(Y_1;V_2|X_1)+ I(Y_2;V_1|X_2)-I(Z;V_1, V_2).\Label{FM_Joint_18}
\end{align} 
Consider \eqref{FM_Joint_6}, \eqref{FM_Joint_13},  \eqref{FM_Joint_15} and \eqref{FM_Joint_18} to remove $R_{2,k}.$ We obtain
\begin{align}
	R_{2,s} + R_{1,e} <& I(Y_1;V_2|X_1)-I(Z;V_2),\Label{FM_Joint_19}\\
	R_{2,s} +R_{2,e}+ R_{1,e} <& I(Y_1;V_2|X_1). \Label{FM_Joint_20}\\
	R_{2,s} + R_{2,e} + R_{1,s} + R_{1,e}  <& I(Y_1;V_2|X_1)+ I(Y_2;V_1|X_2)-I(Z;V_1, V_2).\Label{FM_Joint_21}
\end{align} 
Next, we remove $R_{1,s}$ and $R_{2,s}$ by replacing them by $R_1-R_{1,e}$ and $R_2-R_{2,e}$ (according to \eqref{FM_Joint_8} and \eqref{FM_Joint_9}),  respectively,  in \eqref{FM_Joint_16}, \eqref{FM_Joint_17}, \eqref{FM_Joint_19}, \eqref{FM_Joint_20} and \eqref{FM_Joint_21}. Together with the non-negativity of  $R_{1,s}$ and $R_{2,s}$, we obtain
\allowdisplaybreaks
\begin{align}
	R_1 \geq & R_{1,e}, \Label{FM_Joint_8C}\\ 
	R_2 \geq & R_{2,e},\Label{FM_Joint_9C}\\
	R_1- R_{1,e} + R_{2,e} <& I(Y_2;V_1|X_2)-I(Z;V_1),\Label{FM_Joint_16C}\\
	R_1+ R_{2,e}  <& I(Y_2;V_1|X_2),\Label{FM_Joint_17C}\\
	R_2- R_{2,e} + R_{1,e} <& I(Y_1;V_2|X_1)-I(Z;V_2),\Label{FM_Joint_19C}\\
	R_2+ R_{1,e} <& I(Y_1;V_2|X_1), \Label{FM_Joint_20C}\\
	R_1 + R_2 <& I(Y_1;V_2|X_1)+ I(Y_2;V_1|X_2)-I(Z;V_1, V_2).\Label{FM_Joint_21C}
\end{align}
Consider  \eqref{FM_Joint_8C}, \eqref{FM_Joint_16C},  \eqref{FM_Joint_19C} and \eqref{FM_Joint_20C} to remove $R_{1,e}.$ We obtain
\allowdisplaybreaks
\begin{align}
	R_{2,e} <& I(Y_2;V_1|X_2)-I(Z;V_1), \Label{FM_Joint_16CC}\\
	R_1 +R_2 < & I(Y_1;V_2|X_1)+ I(Y_2;V_1|X_2)-I(Z;V_1)-I(Z;V_2), \Label{FM_Joint_21CC}\\ 
	R_1 +R_2+ R_{2,e}< &  I(Y_1;V_2|X_1)+ I(Y_2;V_1|X_2)-I(Z;V_1),\Label{FM_Joint_22}\\
	R_2- R_{2,e} <& I(Y_1;V_2|X_1)-I(Z;V_2),\Label{FM_Joint_23}\\
	R_2<& I(Y_1;V_2|X_1). \Label{FM_Joint_24}
\end{align}
Consider \eqref{FM_Joint_9C}, \eqref{FM_Joint_17C},  \eqref{FM_Joint_19C}, \eqref{FM_Joint_16CC}, \eqref{FM_Joint_22} and \eqref{FM_Joint_23} to remove $R_{2,e}.$ We obtain
\allowdisplaybreaks
\begin{align}
	R_1 + 2R_2 < & 2I(Y_1;V_2|X_1)+ I(Y_2;V_1|X_2)-I(Z;V_1)-I(Z;V_2), \Label{FM_Joint_24C}\\ 
	R_2<& I(Y_1;V_2|X_1)+ I(Y_2;V_1|X_2)-I(Z;V_1)-I(Z;V_2), \Label{FM_Joint_25}\\
	R_1 +R_2< &  I(Y_1;V_2|X_1)+ I(Y_2;V_1|X_2)-I(Z;V_2),\Label{FM_Joint_26}\\
	R_1 <& I(Y_2;V_1|X_2),\Label{FM_Joint_27}\\
	R_1 +R_2< &  I(Y_1;V_2|X_1)+ I(Y_2;V_1|X_2)-I(Z;V_1),\Label{FM_Joint_28}
\end{align}
and rate conditions that $ I(Y_1;V_2|X_1)\geq I(Z;V_2)$ and $I(Y_2;V_1|X_2)\geq I(Z;V_1).$
Note that \eqref{FM_Joint_24C} is redundant due to \eqref{FM_Joint_21CC} and \eqref{FM_Joint_24}; \eqref{FM_Joint_25}, \eqref{FM_Joint_26} and \eqref{FM_Joint_28} are redundant due to \eqref{FM_Joint_21CC}; and \eqref{FM_Joint_21CC} is redundant due to \eqref{FM_Joint_21C} (as $I(Z;V_1, V_2)\geq I(Z;V_1)+I(Z;V_2)$ since $V_1, V_2$ are independent). So as a summary, we obtain the joint secrecy region of $(R_1, R_2)$ that are constrained by \eqref{FM_Joint_21C}, \eqref{FM_Joint_24} and \eqref{FM_Joint_27}, which is the same as the region defined in \eqref{HH1A}.

\section{Fourier-Motzkin elimination to derive the individual secrecy region}\Label{APP: FM Individual}
To derive the individual secrecy region, recall that we have the following rate constraints:
\allowdisplaybreaks
\begin{align}
	R_1 =& R_{1,s} + R_{1,e}, \Label{FM_Ind_8}\\ 
	R_2 =& R_{2,s} + R_{2,e},\Label{FM_Ind_9}\\
	R_{2,s} + R_{2,k}+R_{2,o} + R_{2,e} <& I(Y_1;V_2|X_1), \Label{FM_Ind_7}\\
	R_{1,s} + R_{1,k}+R_{1,o} + R_{1,e}  <& I(Y_2;V_1|X_2),\Label{FM_Ind_6}\\
	R_{1,e} \leq & R_{2,k}, \Label{FM_Ind_4}\\
	R_{2,e} \leq & R_{1,k}, \Label{FM_Ind_5}\\
	R_{1,o} + R_{1,e} > &I(Z;V_1), \Label{FM_Ind_1}\\
	R_{2,o} + R_{2,e} > & I(Z;V_2),\Label{FM_Ind_2}\\
	R_{1,o} + R_{1,e} +R_{2,o} + R_{2,e}+ \min(R_{1,s}, R_{2,s}) > & I(Z;V_1, V_2). \Label{FM_Ind_3}
\end{align}
First consider \eqref{FM_Ind_6}, \eqref{FM_Ind_1} and \eqref{FM_Ind_3} to remove $R_{1,o}.$ We obtain
\begin{align}
	R_{1,s} + R_{1,k}+ R_{1,e}  <& I(Y_2;V_1|X_2),\Label{FM_Ind_11}\\
	R_{1,s} + R_{1,k}  <& I(Y_2;V_1|X_2)-I(Z;V_1),\Label{FM_Ind_12}\\
	R_{2,o} + R_{2,e}+ \min(R_{1,s}, R_{2,s}) -	R_{1,s} - R_{1,k}> & I(Z;V_1, V_2)- I(Y_2;V_1|X_2). \Label{FM_Ind_13}
\end{align}
Consider \eqref{FM_Ind_7} , \eqref{FM_Ind_2} and \eqref{FM_Ind_13} to remove $R_{2,o}.$ We obtain
\begin{align}
	R_{2,s} + R_{2,k}+R_{2,e} <& I(Y_1;V_2|X_1), \Label{FM_Ind_14}\\
	R_{2,s} + R_{2,k}  <& I(Y_1;V_2|X_1)-I(Z;V_2), \Label{FM_Ind_15}\\
	R_{1,k} + R_{2,k} + \max(R_{1,s}, R_{2,s})  <& I(Y_1;V_2|X_1)+ I(Y_2;V_1|X_2)-I(Z;V_1, V_2).\Label{FM_Ind_16}
\end{align} 
Consider \eqref{FM_Ind_5}, \eqref{FM_Ind_11} and \eqref{FM_Ind_12}  and \eqref{FM_Ind_16} to remove $R_{1,k}.$ We obtain
\begin{align}
	R_{1,s} + R_{1,e}+ R_{2,e}  <& I(Y_2;V_1|X_2),\Label{FM_Ind_17}\\
	R_{1,s} + R_{2,e}  <& I(Y_2;V_1|X_2)-I(Z;V_1),\Label{FM_Ind_18}\\
	R_{2,e} + R_{2,k} + \max(R_{1,s}, R_{2,s})  <& I(Y_1;V_2|X_1)+ I(Y_2;V_1|X_2)-I(Z;V_1, V_2).\Label{FM_Ind_19}
\end{align} 
Consider \eqref{FM_Ind_4}, \eqref{FM_Ind_14} and \eqref{FM_Ind_15}  and \eqref{FM_Ind_19} to remove $R_{2,k}.$ We obtain
\begin{align}
	R_{2,s} + R_{2,e}+R_{1,e} <& I(Y_1;V_2|X_1), \Label{FM_Ind_20}\\
	R_{2,s} + R_{1,e}  <& I(Y_1;V_2|X_1)-I(Z;V_2), \Label{FM_Ind_21}\\
	R_{1,e} + R_{2,e} + \max(R_{1,s}, R_{2,s})  <& I(Y_1;V_2|X_1)+ I(Y_2;V_1|X_2)-I(Z;V_1, V_2).\Label{FM_Ind_22}
\end{align} 
Next, we remove $R_{1,e}$ and $R_{2,e}$ replacing them by $R_1-R_{1,s}$ and $R_2-R_{2,s}$ (according to \eqref{FM_Ind_8} and \eqref{FM_Ind_9}), respectively, in \eqref{FM_Ind_17}, \eqref{FM_Ind_18}, \eqref{FM_Ind_20}, \eqref{FM_Ind_21} and \eqref{FM_Ind_22}. Together with the non-negativity of $R_{1,e}$ and $R_{2,e}$, we obtain
\allowdisplaybreaks
\begin{align}
	R_1 \geq & R_{1,s}, \Label{FM_Ind_8C}\\ 
	R_2 \geq & R_{2,s},\Label{FM_Ind_9C}\\
	R_1+ R_2-R_{2,s}  <& I(Y_2;V_1|X_2),\Label{FM_Ind_17C}\\
	R_{1,s} + R_2-R_{2,s}  <& I(Y_2;V_1|X_2)-I(Z;V_1),\Label{FM_Ind_18C}\\
	R_2 + R_1-R_{1,s} <& I(Y_1;V_2|X_1), \Label{FM_Ind_20C}\\
	R_{2,s} +R_1-R_{1,s}  <& I(Y_1;V_2|X_1)-I(Z;V_2), \Label{FM_Ind_21C}\\
	R_1+ R_2- \min(R_{1,s}, R_{2,s})  <& I(Y_1;V_2|X_1)+ I(Y_2;V_1|X_2)-I(Z;V_1, V_2).\Label{FM_Ind_22C}
\end{align}
Consider  \eqref{FM_Ind_8C}, \eqref{FM_Ind_18C},  \eqref{FM_Ind_20C}, \eqref{FM_Ind_21C} and \eqref{FM_Ind_22C} to remove $R_{1,s}.$ We obtain
\allowdisplaybreaks
\begin{align}
	R_2<& I(Y_1;V_2|X_1), \Label{FM_Ind_23}\\
	R_{2,s} <& I(Y_1;V_2|X_1)-I(Z;V_2), \Label{FM_Ind_24}\\
	R_2 <& I(Y_1;V_2|X_1)+ I(Y_2;V_1|X_2)-I(Z;V_1, V_2),\Label{FM_Ind_25}\\
	R_2-R_{2,s}  <& I(Y_2;V_1|X_2)-I(Z;V_1),\Label{FM_Ind_26}\\
	R_1+2R_2-R_{2,s}  <& I(Y_1;V_2|X_1)+I(Y_2;V_1|X_2)-I(Z;V_1),\Label{FM_Ind_27}\\
	R_1+R_2<& I(Y_1;V_2|X_1)+ I(Y_2;V_1|X_2)-I(Z;V_1)-I(Z; V_2),\Label{FM_Ind_28}\\
	R_1+2R_2-R_{2,s}  <& I(Y_1;V_2|X_1)+2 I(Y_2;V_1|X_2)-I(Z;V_1, V_2)-I(Z;V_1).\Label{FM_Ind_29}
\end{align}
Consider \eqref{FM_Ind_9C}, \eqref{FM_Ind_17C}, \eqref{FM_Ind_22C},  \eqref{FM_Ind_24}, \eqref{FM_Ind_26}, \eqref{FM_Ind_27} and \eqref{FM_Ind_29} to remove $R_{2,s}.$ We obtain
\allowdisplaybreaks
\begin{align}
	R_1 <& I(Y_2;V_1|X_2),\Label{FM_Ind_30}\\
	R_1 +R_2< &  I(Y_1;V_2|X_1)+ I(Y_2;V_1|X_2)-I(Z;V_2),\Label{FM_Ind_31}\\
	R_1 +R_2< &  I(Y_1;V_2|X_1)+ 2I(Y_2;V_1|X_2)-I(Z;V_1, V_2)-I(Z;V_1),\Label{FM_Ind_32}\\
	R_1 +R_2< &  I(Y_1;V_2|X_1)+ I(Y_2;V_1|X_2)-I(Z;V_1),\Label{FM_Ind_33}\\
 	R_2 < & I(Y_1;V_2|X_1)+ I(Y_2;V_1|X_2)-I(Z;V_1)-I(Z;V_2), \Label{FM_Ind_34}\\ 
	R_1+2R_2<&2I(Y_1;V_2|X_1)+ I(Y_2;V_1|X_2)-I(Z;V_1)-I(Z;V_2), \Label{FM_Ind_35}\\
	R_1+2R_2<&2I(Y_1;V_2|X_1)+ 2I(Y_2;V_1|X_2)-I(Z;V_1, V_2)-I(Z;V_1)-I(Z;V_2), \Label{FM_Ind_36}\\
	R_1 <& I(Y_1;V_2|X_1)+ I(Y_2;V_1|X_2)-I(Z;V_1, V_2),\Label{FM_Ind_37}\\
	R_1 +R_2< &  2I(Y_1;V_2|X_1)+ I(Y_2;V_1|X_2)-I(Z;V_1, V_2)-I(Z;V_2), \Label{FM_Ind_38}
\end{align}
and rate conditions that $ I(Y_1;V_2|X_1)\geq I(Z;V_2)$ and $I(Y_2;V_1|X_2)\geq I(Z;V_1).$
Note that \eqref{FM_Ind_31}, \eqref{FM_Ind_33} and \eqref{FM_Ind_34} are redundant due to \eqref{FM_Ind_28}; \eqref{FM_Ind_35} is redundant due to \eqref{FM_Ind_28} and \eqref{FM_Ind_23}; and \eqref{FM_Ind_36} is redundant due to \eqref{FM_Ind_25} and \eqref{FM_Ind_28}. So as a summary, we obtain the individual secrecy region of $(R_1, R_2)$ that are constrained by \eqref{FM_Ind_23}, \eqref{FM_Ind_25}, \eqref{FM_Ind_28},  \eqref{FM_Ind_30}, \eqref{FM_Ind_32}, \eqref{FM_Ind_37} and \eqref{FM_Ind_38}, which is the same as the region defined in \eqref{HH2A}.

\end{document}